\documentclass[a4paper,11pt]{article}
\pdfoutput=1

\usepackage{jinstpub} 
\usepackage{textcomp}

\usepackage{amsmath, amsthm, amssymb}
\usepackage[ansinew]{inputenc}
\usepackage{wrapfig}
\usepackage{graphicx}
\usepackage{url}
\usepackage{float}

\usepackage{xspace}

\mathchardef\mhyphen="2D
\title{In-situ calibration of the single-photoelectron charge response of the IceCube photomultiplier tubes}

\collaboration{\includegraphics[height=17mm]{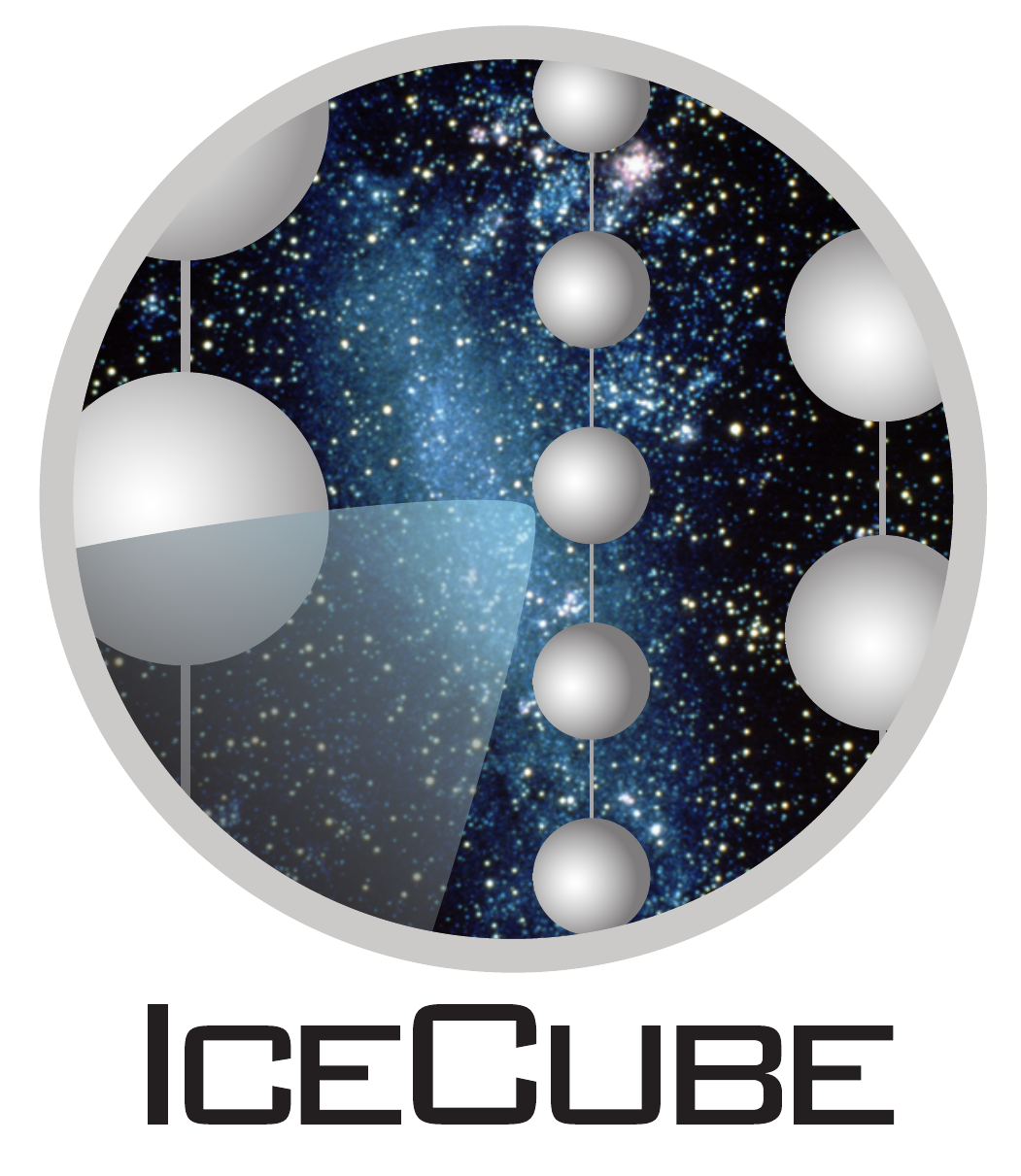}\\[6pt]IceCube collaboration}
\author[p]{M.~G.~Aartsen,}
\author[bc]{M.~Ackermann,}
\author[p]{J.~Adams,}
\author[l]{J.~A.~Aguilar,}
\author[t]{M.~Ahlers,}
\author[at]{M.~Ahrens,}
\author[z]{C.~Alispach,}
\author[ak]{K.~Andeen,}
\author[az]{T.~Anderson,}
\author[l]{I.~Ansseau,}
% ...
\affiliation[l]{Universit{\'e} Libre de Bruxelles, Science Faculty CP230, B-1050 Brussels, Belgium}
% ...
\affiliation[p]{Dept. of Physics and Astronomy, University of Canterbury, Private Bag 4800, Christchurch, New Zealand}
% ...
\affiliation[t]{Niels Bohr Institute, University of Copenhagen, DK-2100 Copenhagen, Denmark}
% ...
\affiliation[z]{D{\'e}partement de physique nucl{\'e}aire et corpusculaire, Universit{\'e} de Gen{\`e}ve, CH-1211 Gen{\`e}ve, Switzerland}
% ...
\affiliation[ak]{Department of Physics, Marquette University, Milwaukee, WI, 53201, USA}
% ...
\affiliation[at]{Oskar Klein Centre and Dept. of Physics, Stockholm University, SE-10691 Stockholm, Sweden}
% ...
\affiliation[az]{Dept. of Physics, Pennsylvania State University, University Park, PA 16802, USA}
% ...
\affiliation[bc]{DESY, D-15738 Zeuthen, Germany}
% ...

\emailAdd{analysis@icecube.wisc.edu}

\abstract{We describe an improved in-situ calibration of the single-photoelectron charge distributions for each of the in-ice Hamamatsu Photonics R7081-02[MOD] photomultiplier tubes in the IceCube Neutrino Observatory. The characterization of the individual PMT charge distributions is important for PMT calibration, data and Monte Carlo simulation agreement, and understanding the effect of hardware differences within the detector. We discuss the single photoelectron identification procedure and how we extract the single-photoelectron charge distribution using a deconvolution of the multiple-photoelectron charge distribution. 
}


\arxivnumber{tbd} 

\begin{document}
\maketitle
\flushbottom

\section{Introduction}\label{sec::introduction}

This article discusses a method for determining the in-situ single-photoelectron (SPE) charge distributions of individual photomultiplier tubes~(PMT) used in the IceCube neutrino detector. The SPE charge distribution refers to the charge probability density function of an individual PMT generated by the amplification of a pure sample of single photoelectrons. An accurate measurement of the individual SPE charge distribution is important for PMT calibration, since a change in the PMT gain corresponds to a proportional change in the extracted charge. Beyond this, a more accurate description of the individual SPE charge distributions in Monte Carlo (MC) simulation can also improve the overall data/MC agreement. The features extracted from the SPE charge distributions are also useful for examining hardware difference, such as the quantum efficiency, and assessing long term stability of the detector. 

\newpage
\section{Experimental setup}\label{sec::setup}

\begin{wrapfigure}[20]{R}{0.6\textwidth}
\centering
\includegraphics[width=.6\textwidth]{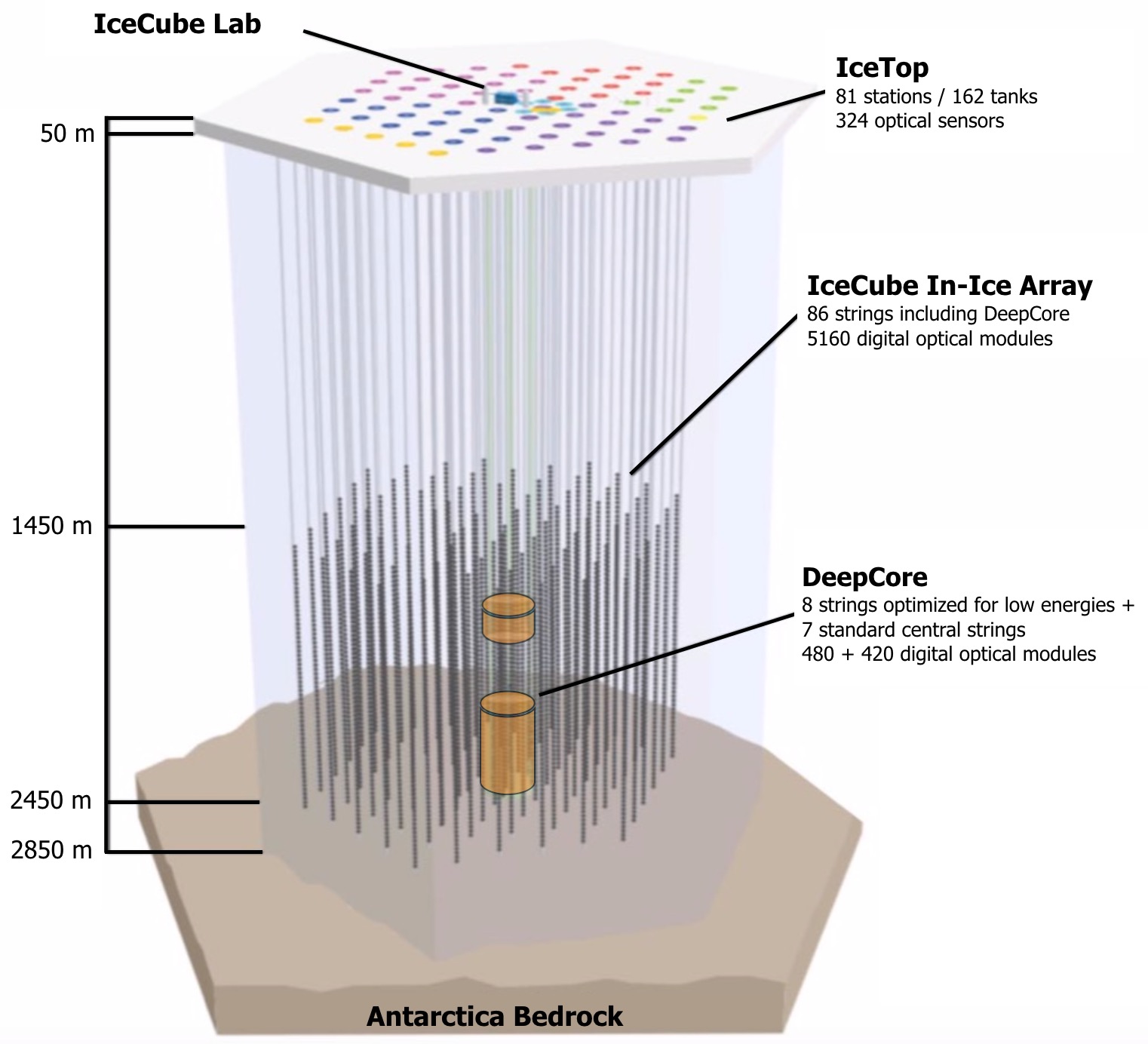}
\caption{The IceCube Neutrino Observatory. From Ref.~\cite{halzen2010invited}.}
\label{fig::ICL}
\end{wrapfigure}

The IceCube Neutrino Observatory~\cite{icecube_design,achterberg2006first} (Fig.~\ref{fig::ICL}) is a cubic-kilometer-sized array of 5,160 photomultiplier tubes buried in the Antarctic ice sheet, designed to observe high-energy neutrinos interacting with the ice~\cite{icecube_astro}. In 2011, the IceCube Collaboration completed the installation of 86 vertical \textit{strings} of PMT modules, eight of which were arranged in a denser configuration known as the DeepCore sub-array~\cite{deepcore}. Each string in IceCube contains 60 digital optical modules (DOMs), which contain a single PMT each, as well as all required electronics~\cite{abbasi2009icecube}. The primary 78 strings (excluding DeepCore) are spaced 125\,m apart in a hexagonal grid, with the DOMs extending from 1450\,m to 2450\,m below the surface of the ice sheet. The additional DeepCore strings (79-86) are positioned between the centermost strings in the detector, reducing the horizontal DOM-to-DOM distance in this region to between 42\,m and 72\,m. The lower 50 DOMs on these strings are located in the deepest 350\,m of the detector surrounded by the cleanest ice~\cite{aartsen2013measurement}, while the upper ten provide a cosmic ray veto extending down from 1900\,m to 2000\,m below the surface. Above the in-ice detectors, there exists a surface array, IceTop~\cite{abbasi2013icetop}, consisting of 81 stations located just above the in-ice IceCube strings. The PMTs located in IceTop DOMs operate at a lower gain  and the data from these PMTs was not included in the current analysis; however, the IceTop PMTs are calibrated to single photoelectron charge distribution in a similar way as the in-ice PMTs (see Sec. 5.1 in Ref.~\cite{abbasi2013icetop}).

%

Each DOM consists of a 0.5"-thick spherical glass pressure vessel that houses a single down-facing 10" PMT from Hamamatsu Photonics. The PMT is coupled to the glass housing with optical gel and is surrounded by a wire mesh to reduce the effect of the Earth's ambient magnetic field. The glass housing is transparent to wavelengths of 350\,nm and above~\cite{icecube_pmt}. 

\begin{figure}[t!]
    \centering
    \begin{minipage}{.5\textwidth}
        \centering
        \includegraphics[width=1\linewidth]{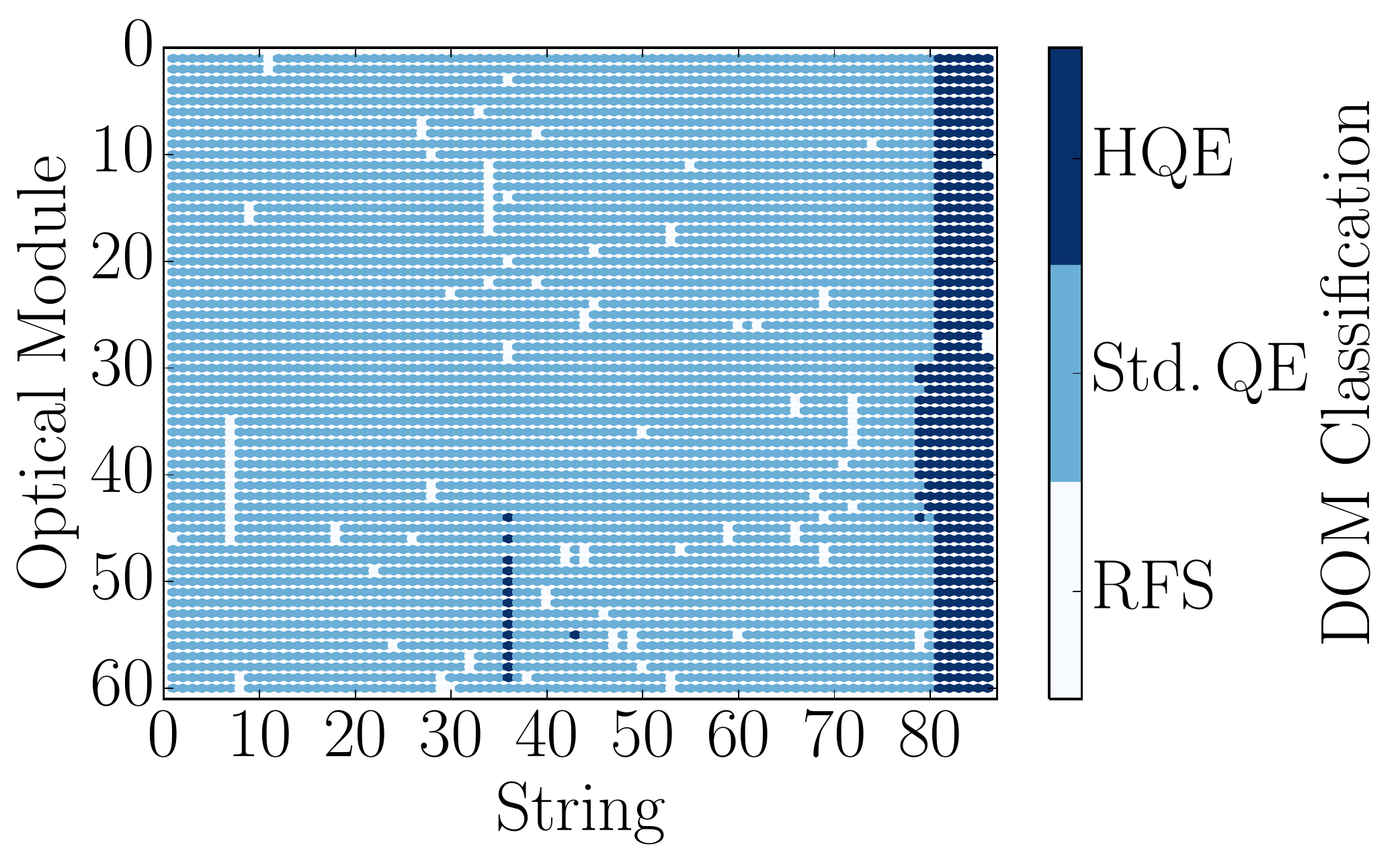}
    \end{minipage}%
    \begin{minipage}{0.5\textwidth}
        \centering
        \includegraphics[width=1\linewidth]{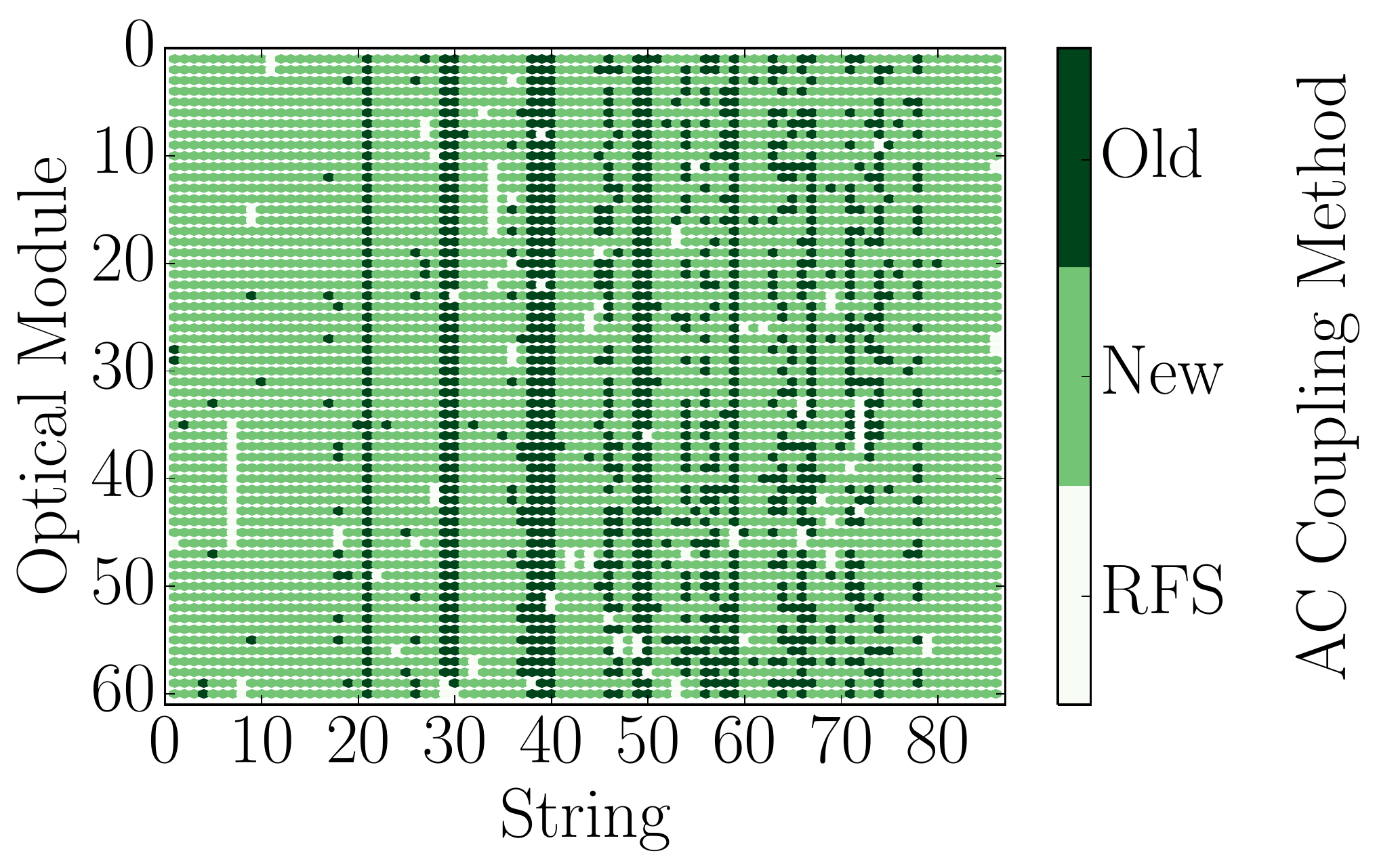}
    \end{minipage}
    \caption{Left: A mapping of the HQE (dark blue) and Standard QE DOMs (light blue). Right: The version of AC coupling, old toroids (dark green) and new toroids (light green). DOMs that were removed from service (RFS) for the IC86.2016 season are shown in white. These are primarily DOMs that resulted in failures during deployment and failures during subsequent operation. Due to subsequent DOM failures, the number of RFS DOMs from IC86.2011 to IC86.2016 varies between 107 to 111.}
	\label{fig::DOM_features}
\end{figure}

Of the 5,160 DOMs, 4,762 house a R7081-02 Hamamatsu Photonics PMT, sensitive to wavelengths ranging from 300\,nm to 650\,nm, with a peak quantum efficiency of 25\% near 390\,nm. These are classified as Standard Quantum Efficiency (Standard QE) DOMs. The remaining 398 DOMs are equipped with the Hamamatsu R7081-02MOD PMTs, which, having a peak quantum efficiency of 34\% near 390\,nm (36\% higher efficiency than the Standard QE DOMs), are classified as High Quantum Efficiency (HQE) DOMs~\cite{deepcore}. These DOMs are primarily located in DeepCore and on strings 36 and 43, as shown in the left side of Fig.~\ref{fig::DOM_features}. 

The R7081-02 and R7081-02MOD PMTs have 10 dynode stages and are operated with a nominal gain of 10$^7$ and achieved with high voltages ranging from approximately 1215$\,\pm$\,83\,V and 1309\,$\pm$\,72\,V, respectively. A typical amplified single photoelectron generates a 5.2\,$\pm$\,0.3\,mV peak voltage after digitization with a full width at half maximum of 13\,$\pm$\,1\,ns. The PMTs are operated with the anodes at high voltage, so in order to remove the DC voltage from the signal, the PMT is AC coupled to the amplifiers (front-end amplifiers). There are two versions of AC coupling employed in the PMTs, referred to as the \textit{new} and \textit{old toroids}, both of which use custom-designed toroidal transformers. The new toroids were designed with a larger ferrite core and more turns resulting in a droop time constant of 15$\mu$s, ten times larger than the time constant of the old toroids. The DOMs instrumented with the old toroids were also designed with an impedance of 43\,$\Omega$, while the new toroids are 50\,$\Omega$. The toroidal transformer effectively acts as a high-pass filter with an approximately flat frequency response up to 100MHz. Single photoelectrons, measured at the output of the secondary winding, however, show percent-level shape differences between new and old toroids. The toroidal coupling method was chosen partially since it offers a higher level of reliability than capacitive coupling and has lower stored energy for the same frequency response~\cite{icecube_instrumentation}. Conventional AC-coupling high-voltage ceramic capacitors can also produce undesirable noise from leakage currents and are impractical given the signal droop and undershoot requirements~\cite{icecube_pmt}.  The locations of DOMs with the different versions of AC-coupling are shown on the right side of Fig.~\ref{fig::DOM_features}.  All HQE DOMs are instrumented with the new toroids. 

IceCube relies on two observables per DOM to reconstruct events: the total number of detected photons and their timing distribution. Both the timing and the number of photons are extracted from the digitized waveforms. This is accomplished by deconvolving the digitized waveforms into a superimposition of single photoelectron pulses (so-called pulse series), and the integral of the individual pulses divided by the load resistance defines the observed charge. It will often be expressed in units of PEs, or photoelectrons, which further divides the measured charge by the charge of a single electron times the nominal gain. 

When one or more photoelectrons produce a voltage at the anode sufficient to trigger the onboard discriminator, the signal acquisition process is triggered. The discriminator threshold is set to approximately 1.2\,mV, or equivalently to $\sim$0.23\,PE, via a digital-to-analog converter (DAC). The signal is presented to four parallel channels for analog amplification and digitization. The highest-gain channel has a nominal amplification of 16 and is most suitable for single photon detection, and is thus used in this analysis. This channel is digitized at 300\,million samples per second (MSPS) for 128 samples (the readout time window of 427\,ns) using a 10-bit Analog Transient Waveform Digitizer (ATWD). 
Two ATWD chips are present on the DOM Mainboard (MB) and alternate digitization between waveforms to remove dead time associated with the readout. 




\subsection{Single-photoelectron charge distributions} \label{sec::SPE_Templates}
The assessment of the in-situ single-photoelectron charge distributions was made possible with the development of two pieces of software:

\begin{enumerate}
\item A specially-designed unbiased pulse selection developed to reduce the multiple photoelectron (MPE) contamination while accounting for other physical phenomena (e.g.~late pulses, afterpulses, pre-pulses, and waveform distortions) and software-related effects (e.g.~pulse splitting). This is further described in Sec.~\ref{sec::pulse_selection}.
\item A fitting procedure developed to separate the remaining MPE contamination from the SPE charge distribution by deconvolving the measured charged distribution. This is further described in Sec.~\ref{sec::fitting}.
\end{enumerate}

By using in-situ data to determine the SPE charge distributions, we accurately represent the individual PMT response as a function of time, environmental conditions, software version and hardware differences, and realistic photocathode illumination conditions. This is beneficial since it also allows us to inspect the stability and long-term behavior of the individual DOMs, verify previous calibration, and correlate features with specific DOM hardware.

Ideally, a single photon produces a single photoelectron, which is then amplified by a known amount, and the measured charge is calibrated to correspond to 1\,PE. However, there are many physical processes that create a spurious structure in the measured charge distributions. For example:

\begin{itemize}
\item \textbf{Statistical fluctuation due to cascade multiplication}~\cite{hamamatsu_pmt}. At every stage of dynode amplification, the number of emitted electrons that make it to the next dynode is randomly distributed. This in turn causes a smearing in the measured charge after the gain stage of the PMT. 

\item \textbf{Photoelectron trajectory}. Some electrons may deviate from the favorable trajectory, reducing the number of secondaries produced at a dynode or the efficiency to collect them on the following dynode. This can occur at any stage, but it has the largest effect on the multiplication at the first dynode~\cite{hamamatsu_handbook}. The trajectory of a photoelectron striking the first dynode will depend on many things, including where on the photocathode it was emitted, the uniformity of the electric field, the size and shape of the dynodes~\cite{hamamatsu_pmt}, and the ambient magnetic field~\cite{PMTMagnetic,doublechooz_magnetic}. 

\item \textbf{Late or delayed pulses}. A photoelectron can elastically or inelastically backscatter off the first dynode. The scattered electron can then be re-accelerated to the dynode, creating a second pulse. The difference in time between the initial pulse and the re-accelerated pulse in the R7081-02 PMT was previously measured to be up to 70~ns~\cite{icecube_pmt,charge_distribution}. Elastically backscattered photoelectrons will carry the full energy and are thus expected to produce similar charge to a non-backscattered photoelectron, albeit with a time offset. The mean measured charge of an ineleastic backscattered photoelectron, by contrast, is expected to be smaller than a nominal photoelectron~\cite{lubsandorzhiev2000studies}.

\item  \textbf{Afterpulses}. When photoelectrons or the secondary electrons produced during the electron cascade gain sufficient energy to ionize residual gas in the PMT, the resulting positively charged ionized gas will be accelerated in the electric field towards the photocathode. Upon impact with the photocathode, multiple electrons can be released from the photocathode, creating what is called an afterpulse. For the R7081-02 PMTs used in IceCube, the timescale for afterpulses was measured to occur from 0.3 to 11\,$\mu$s after the initial pulse, with the first prominent afterpulse peak occurring at approximately 600~ns~\cite{icecube_pmt}. The spread in the afterpulse time depends on the position of photocathode, the charge-to-mass ratio of the ion produced, and the electric potential distribution~\cite{ma2011time}, whereas the size of the afterpulse is related to the momentum and species of the ionized gas and composition of the photocathode~\cite{torre1983study}. 

\item \textbf{Pre-pulses}. If an incident photon passes through the photocathode without interaction and strikes the first dynode, it can eject an electron that is only amplified by the subsequent stages, resulting in a lower measured charge (lower than the SPE pulse size by a factor of 10 to 20).  For the IceCube PMTs, the prepulses have been found to arrive approximately 30\,ns before the signal from photoelectrons emitted from the photocathode. The rate of pre-pulses was measured to be less than 1\% of the SPE rate~\cite{icecube_pmt}. 

\item \textbf{MPE contamination}. When multiple photoelectrons arrive at the first dynodes within few nanoseconds of each other, they can be reconstructed by the software as a single MPE pulse.

\item \textbf{Dark noise}. Photoelectron emission, not initiated from an external event, can be attributed to thermionic emission from the low work function photocathode and the dynodes, Cherenkov radiations initiated from radioactive decay within the DOM, and field emission from the electrodes. Dark noise originating from thermionic emission from the dynodes is shown in Ref.~\cite{hamamatsu_construction} to populate the low-charge region. In the temperature range of interest for IceCube (-40$^\circ$ to -20$^\circ$) the dark noise rate for the R7081-02 PMTs are approximately 350~Hz~\cite{icecube_pmt}.



\item \textbf{Electronic noise}. This refers to the combined fluctuations caused by noise generated from the analog-frontend and the analog-to-digital converters (ATWDs and fADC). When integrated over a time window the resulting charge is generally small and centered around zero, thus only leading to a small broadening in the low charge region. The standard deviation of the electronic noise was found to be approximately $\pm$0.11~mV.
\end{itemize}

Beyond the physical phenomena above that modify the measured charge distribution, there is also a lower limit on the smallest charge that can be extracted. For IceCube, the discriminator only triggers for peak voltages above the threshold and subsequent pulses in the readout window are subject to a threshold defined in the software. This software threshold was set conservatively to avoid extracting pulses that originated from electronic noise. It can be modified to gain access to lower charge pulses and will be discussed in Sec.~\ref{sec::lowPE} at the expense of introducing more electronic noise into the charge distribution.

The standard SPE charge distribution used for all DOMs in IceCube, known as the TA0003 distribution~\cite{icecube_pmt}, models the above effects as the sum of an exponential plus a Gaussian. The TA0003 distribution is the average SPE charge distribution extracted from a laboratory measurement of 118 Hamamatsu R7081-02 PMTs. The measurement was performed in a -32$^\circ$C freezer using a pulsed UV LED centered along the axis of the PMT, directly in front of the photocathode. 

In 2013, IceCube has made several laboratory measurements of the SPE charge distribution of R7081-02 PMTs using single photons generated from synchronized short duration laser pulses. The coincident charge distribution generated by the laser pulses was found to include a steeply falling low-charge component in the region below the discriminator threshold. To improve the fit quality and account for this, a new functional form including a second exponential was introduced. This form of the normalized charge probability distribution $f(q)_{\mathrm{SPE}}=$ Exp$_1$ + Exp$_2$ + Gaussian, is referred to as the \textit{SPE charge template} in this article. Explicitly, it is:

\begin{equation}\label{eq::fit}
f(q)_{\mathrm{SPE}} = \frac{\mathrm{P}_\mathrm{e1}}{\mathrm{w}_1} \cdot  e^{-q/\mathrm{w}_1} +  \frac{\mathrm{P}_\mathrm{e2}}{\mathrm{w}_2} \cdot  e^{-q/\mathrm{w}_2} + \frac{\mathrm{1-\mathrm{P}_\mathrm{e1} -\mathrm{P}_\mathrm{e2}}}{\sigma \sqrt{\pi/2} \cdot \mathrm{Erfc[-\mu/(\sigma \sqrt{2})]}} \cdot  e^{-\frac{(q-\mu)^2}{2\sigma^2}},
\end{equation}
where $q$ represents the measured charge; w$_1$ and w$_2$ are the exponential decay widths; and $\mu$, $\sigma$ are the Gaussian mean and standard deviation (SD), respectively. The coefficients P$_{\mathrm{e}1}$, P$_{\mathrm{e}2}$, and $1-\mathrm{P}_{\mathrm{e}1}-\mathrm{P}_{\mathrm{e}2}$ correspond to the probability of a photoelectron contributing to each component of the SPE template. We will subsequently refer to the probability of a photoelectron contributing to the Gaussian component as $\mathrm{N} =1-\mathrm{P}_{\mathrm{e}1}-\mathrm{P}_{\mathrm{e}2}$. 
The Erfc function used to normalize the Gaussian represents the complementary error function defined as $\mathrm{Erfc}[z]=\frac{2}{\sqrt{\pi}}\int_{z}^{\infty} \mathrm{e}^{-x^2} dx $. Eq.~\ref{eq::fit} is the assumed functional shape of the SPE charge distributions, and the components of Eq.~\ref{eq::fit} are determined in this article for all in-ice DOMs. IceCube has chosen to define 1\,PE as the location of the Gaussian mean ($\mu$) and calibrates the gain of the individual PMTs prior to the start of each season to meet this definition. Any overall bias in the total observed charge can be absorbed into an efficiency term, such as the quantum efficiency. This is valid since the linearity between the instantaneous total charge collected and the number of incident photons is satisfied up to $\sim$2\,V~\cite{icecube_instrumentation}, or approximately 560\,SPEs. That is, the average charge collected from $n$ photons is $n$ times the average charge of the SPE charge distribution, and the average charge of the SPE charge distribution is always a set fraction of the Gaussian mean.


\subsection{IceCube datasets and software definitions}\label{sec::datasets}
The amount of observed light depends on the local properties of the ice~\cite{aartsen2013measurement}. Short term climate variations from volcanoes and longer-term variations from atmospheric dust affect the optical properties of the ice, producing nearly horizontal layers.  This layered structure affects how much light the DOMs observe, and, with it, the trigger rate. The largest contribution to the IceCube trigger rate comes from downward-going muons produced in cosmic ray-induced showers~\cite{icecube_cr}. Cosmic ray muons stopping in the detector cause the individual trigger rates to decrease at lower depths. 





If a DOM and its nearest or next-to-nearest neighbor observe a discriminator threshold crossing within a set time window, a \textit{Hard Local Coincidence} (HLC) is initiated, and the corresponding analog waveforms are sampled and read out on the ATWD channels. Thermionic emission induced dark noise can be present in the readout, however it is suppressed at lower temperatures and is unlikely to trigger an HLC event. 



After waveform digitization, there is a correction applied to remove the DC baseline offset. Distortions to the waveform, such as from droop and undershoot (see Ref.~\cite{leo2012techniques} for a technical description) introduced by the AC coupling, are compensated for in software during waveform calibration. The area encompassed by the distortion of the waveform due to these effects is proportionally related to the charge that produced them, and is therefore commonly a problem for bright events or high frequency trains of photons. The expected distortion can be calculated if the input charge is known, and is compensated for by adding the determined reaction voltage of the distortion to the calibrated waveform. If the undershoot voltage drops below 0~ADC counts (outside of the minimum ADC range), the ADC values are zeroed and then compensated for once the waveform is above the minimum ADC input.


The digital waveforms are then deconvolved into a pulse series using software called "WaveDeform" (waveform unfolding process)~\cite{wavedeform}. The pulse series for a particular DOM corresponds to a list of single photoelectron charges and timestamps. WaveDeform is a fitting algorithm that determines the pulse series by fitting a superposition of single photoelectron basis functions to the the digitized waveforms. The basis functions, or SPE pulse templates, depend on the version of the AC-coupling instrumented in the particular DOM. The pulse series, extracted from various datasets, represents the fundamental unit of data used in this analysis. 




The pulse series used in this analysis come from two datasets:
\begin{enumerate}
\item The \textbf{Event dataset.} This dataset preserves the full digital waveform readout of randomly-selected HLC events, collecting on average 1 in every 1000 HLC events. The largest contribution to this dataset comes from downward-going muons produced in cosmic-ray-induced showers. The average event consists of approximately 26\,PE  distributed over an average of 16 DOMs that observed a discriminator crossing. The full digital waveform of these events allows us to extract the raw information about the individual pulses. This dataset will be used to measure the individual PMT charge distributions. 

\item The \textbf{Forced-Triggered (FT) dataset.} This dataset is populated with digitized waveforms that are initiated by the electronics (forced-triggered) of a channel that has not gone above the threshold. The forced triggered waveforms are typically used to monitor the individual DOM baselines and thus includes the full ATWD waveform readout. Since this dataset is forced-triggered, the majority of these waveforms represent electronic noise with minimal contamination from random accidental coincidence SPEs. This dataset will be used to examine the noise contribution to the charge distributions.  


\end{enumerate}

This analysis uses the full Event and FT datasets from IceCube seasons 2011 to 2016~\cite{aartsen2019search}, subsequently referred to as IC86.2011 to IC86.2016. Seasons in IceCube typically start in May of the labeled year and end approximately one year later. Calibration is performed before the start of each season.

\section{Extracting the SPE charge templates}
\subsection{Single photoelectron pulse selection}\label{sec::pulse_selection}
\begin{figure}[b!]
    \centering
    \begin{minipage}{.48\textwidth}
        \centering
        \includegraphics[width=1\linewidth]{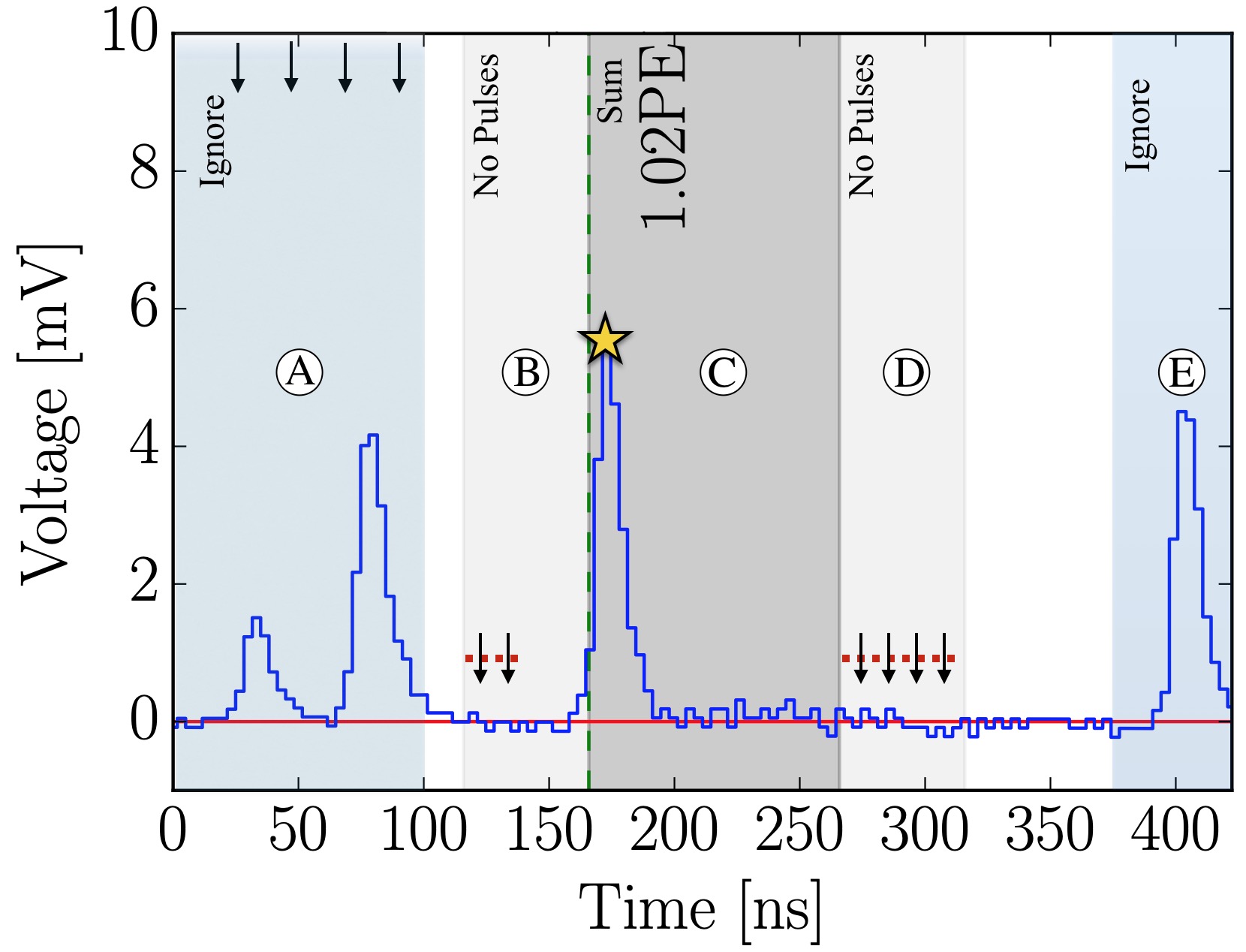}
    \end{minipage}%
    \begin{minipage}{0.48\textwidth}
        \centering
        \includegraphics[width=1\linewidth]{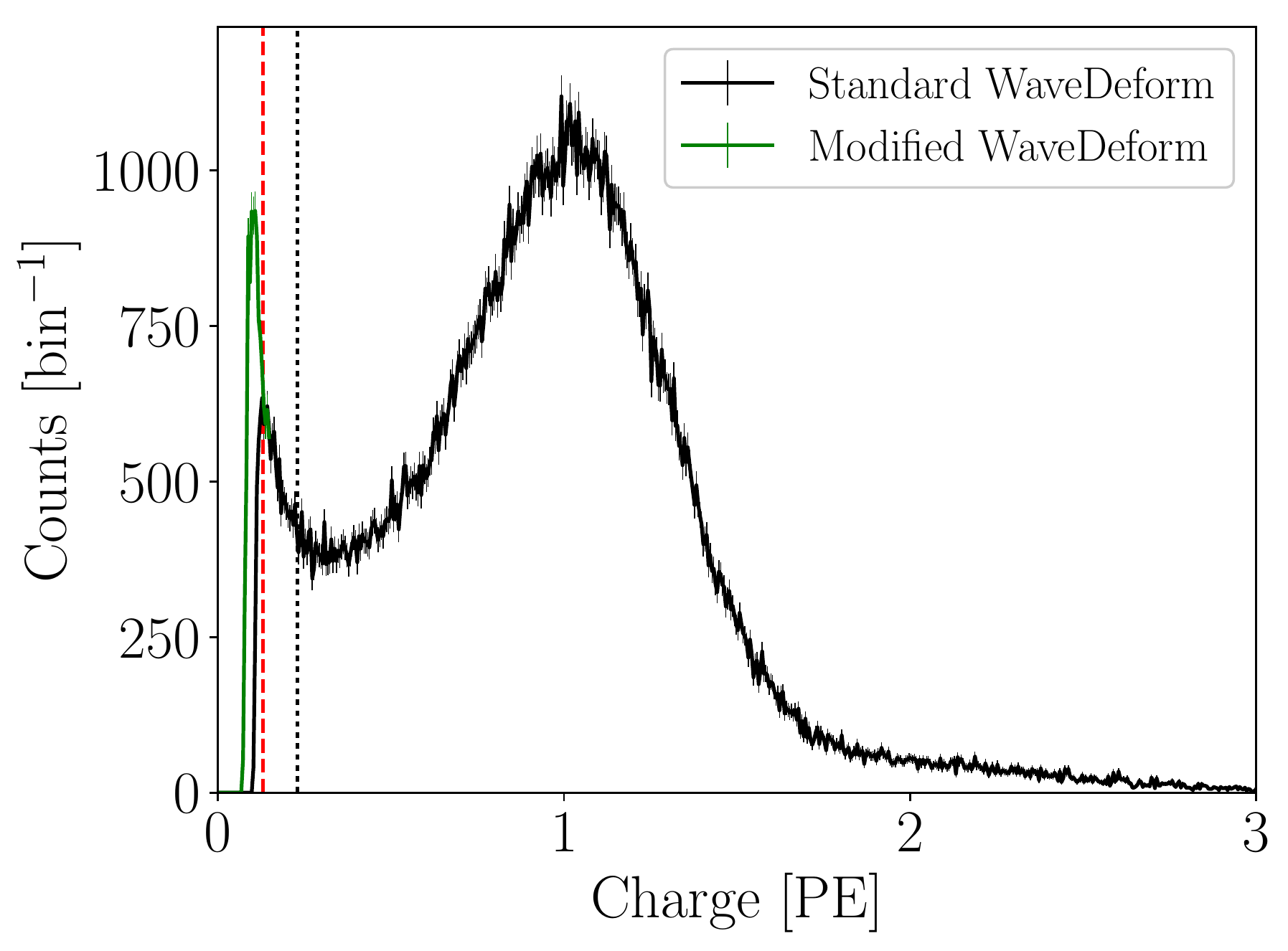}
    \end{minipage}
    \caption{Left: An illustrative diagram of the pulse selection criteria for selecting a high-purity and unbiased sample of single photoelectrons. An example digitized ATWD waveform of data is shown in blue and the baseline is shown as a solid red line. The pulse of interest is identified with a yellow star. This example waveform was triggered by a small pulse at 25\,ns (recall that the delay board allows us to examine the waveform just prior to the trigger pulse), followed by a potential late pulse at 70\,ns. At 400\,ns, we see a pulse in the region susceptible to afterpulses. Waveform voltage checks are illustrated with arrows, and various time windows described in the text are drawn with semi-opaque regions. The pulse of interest (POI) is reported to have a charge of 1.02\,PE, given by WaveDeform, and would pass the pulse selection criteria. Right: The measured charge distribution resulting from the pulse selection for string 1, optical module 1 (DOM 1,1), from the Event dataset (black data points with statistical error bars). As described in the text, the pulse selection is able to extract charge below the discriminator threshold of 0.23\,PE (black dotted line). The threshold shown in the black data points below 0.13\,PE (red dashed line) is due to the termination condition in WaveDeform. After setting to termination condition to a lower value (modified WaveDeform), the charge distribution is extended to lower charges, as exemplified by the green data points.}
	\label{fig::pulse_selection}
\end{figure}

The pulse selection is the method used to extract candidate, unbiased, single photoelectron pulses from highest-gain ATWD channel while minimizing the MPE contamination. The pulse selection was designed such that it avoids collecting afterpulses, does not include late pulses originating from the trigger pulse, accounts for the discriminator threshold, reduces the effect of signal droop and undershoot, and provides sufficient statistics to perform a season-to-season measurement. An illustrative diagram of the pulse selection is shown in the left side of Fig.~\ref{fig::pulse_selection}, while a description of the procedure is detailed below.

We restrict the pulse selection to only extract information from waveforms in which the trigger pulse does not exceed 10\,mV ($\sim$2\,PE) and no subsequent part of the waveform exceeds 20\,mV ($\sim$4\,PE). Here, since we are placing a limit on the input charge, this reduces the effect of the baseline undershoot due to the AC coupling or other artifacts from large pulses. 

In order to trigger a DOM, the input to the front-end amplifiers must exceed the discriminator threshold. To avoid the selection bias of the discriminator trigger (i.e. only selecting pulses greater than the discriminator threshold), we ignore the trigger pulse as well as the entire first 100\,ns of the time window. This is shown as region A in Fig.~\ref{fig::pulse_selection} (left). Ignoring the first 100\,ns removes late pulses that could be attributed to the triggering pulse, which occurs approximately 4\% of the time~\cite{icecube_pmt}. 

To ensure we are not accepting afterpulses into the selection (recall that the first prominent afterpulse peak occurs at approximately 600\,ns), we also enforce the constraint that the pulse of interest (POI) is within the first 375\,ns of the ATWD readout time window. This also allows us to examine the waveform up to 50\,ns after the POI (the full ATWD readout window is 427\,ns). In the vicinity of the POI, we ensure that WaveDeform did not reconstruct any pulses up to 50\,ns prior to the POI, or 100 to 150\,ns after the POI (these regions are labelled as B and D in Fig.~\ref{fig::pulse_selection} (left)). This latter constraint is to reduce the probability of accidentally splitting a late pulse in the summation window.

If a POI is reconstructed between 100 and 375\,ns after the start of the waveform and the voltage criteria above are met, it is accepted as a candidate photoelectron and several checks are performed on the waveform prior to and after the POI.  The first check is to ensure that the waveform is near the baseline just before the rising edge of the POI.  This is accomplished by ensuring that the waveform does not exceed 1\,mV, 50 to 20\,ns prior to the POI, and eliminates cases where the POI is a late pulse. We also ensure the waveform returns to the baseline by checking that no ADC measurement exceeds 1\,mV, 100 to 150\,ns after the POI. These constraints are illustrated as the horizontal red dotted lines and black arrows in regions B and D in the left side of Fig.~\ref{fig::pulse_selection}. 

If all the above criteria are met, we sum the reconstructed charges in the pulse series from the POI time, given by WaveDeform, to +100\,ns (region C in Fig.~\ref{fig::pulse_selection} (left)). This ensures that all nearby pulses are either fully separated or fully added. This summation is critical, since WaveDeform may occasionally split a true SPE pulse into multiple smaller pulses, therefore we must always perform a summation of the charge within a time window to produce a meaningful charge distribution. Consequently, the 100\,ns summation (region C) also implies that the pulse selection will occasionally accept MPE events. We chose 100\,ns window for the summation to be sufficiently large that we collect the charge of a late pulse (recall that late pulses were measured up to 70\,ns after the main pulse), should it be there, while minimizing the MPE contamination. After running the pulse selection, we estimate that there is on average a 6.5\% probability of the summation time window includes two or more photons. 

\subsection{Characterizing the low-charge region}\label{sec::lowPE}
This analysis aims to describe the entire charge probability distribution down to small charge values for each DOM. This is required by the IceCube simulation. However, we cannot extract charge to arbitrary low values before electronic noise begins to dominate. Fig.~\ref{fig::pulse_selection} (right) shows the charge distributions from the Event dataset as black data points after the pulse selection for string 1, optical module 1, DOM(1,1). By ignoring the first 100\,ns of the ATWD time window, we see that we are capable of extracting charge below the discriminator threshold (vertical black dashed line). 


The threshold observed below 0.13\,PE is due to the software threshold mentioned in Sec.~\ref{sec::SPE_Templates}. The threshold is a result of the settings employed within WaveDeform. These settings include a termination condition on the SPE pulse template fitting algorithm that monitors the gradient of the least-squares residual in the fit to the digitized waveforms. When the improvement to the fit, per PE, of allowing one more non-zero pulse drops below the value of the termination condition, the algorithm will terminate. By reducing the value of the termination condition setting, WaveDeform is slower but will produce a better fit to the data and fit and include smaller pulses. The extension to the charge distribution is shown as the green data points in Fig.~\ref{fig::pulse_selection} (right). Reducing this value, however, will also increase the electronic noise contamination. An assessment of this follows.

The steeply falling component above the WaveDeform threshold and below the discriminator (0.13\,PE to 0.23\,PE) is in agreement with the laser measurements mentioned in Sec.~\ref{sec::SPE_Templates}. Since small pulses below the DOM's discriminator threshold will be recorded in events with
multiple photoelectrons, it is important that the SPE charge distribution accounts for this shape. Exp$_1$ was specifically introduced into Eq.~\ref{eq::fit} to account for this component. It was an empirical choice to use an exponential to represent this region of the charge distribution.




\begin{figure}[!h]
\begin{center}
\includegraphics[width=1.\columnwidth]{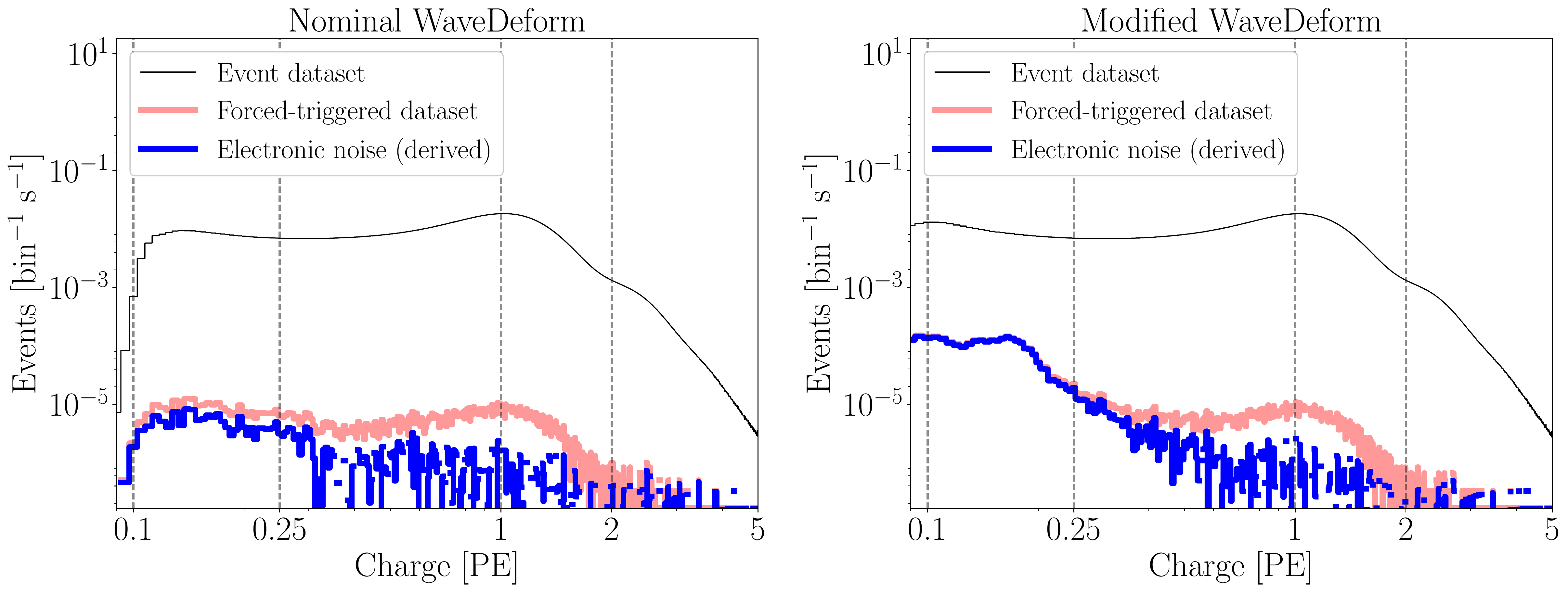}
\caption{The charge distributions of all DOMs for the Event (E) and Forced-triggered (FT) datasets summed over all seasons. The blue histogram shows the derived contribution from electronic noise. This was found by subtracting the Event dataset (normalized such that the rate between E and FT are the same 1PE) from the FT dataset: (FT - E$\times$(FT$|_{1\mathrm{PE}}$/E$|_{1\mathrm{PE}}$)). Left: The charge distributions for the standard WaveDeform settings. Right: The charge distributions for the modified WaveDeform settings.}
  \label{fig::beacon_launches}
 \end{center}
\end{figure}

Fig.~\ref{fig::beacon_launches} (left) shows the charge distributions summed over all DOMs for the Event (black) and the FT (red) datasets using the default settings of WaveDeform (standard WaveDeform). As mentioned in Sec.~\ref{sec::datasets}, occasionally a photoelectron will be coincident with the FT time window. These charges populate an SPE charge distribution. Subtracting the shape of the Event dataset charge distribution -- which is dominated by single photoelectrons -- from the FT dataset, yields an estimate of the amount of electronic noise contamination (blue). The signal-to-noise ratio (SNR) above 0.1~PE is found to be SNR\,$=744.7$ and SNR$\,=\,1.98\times10^5$, in the bin with the lowest SNR value and for the full distribution, respectively.


Fig.~\ref{fig::beacon_launches} (right) shows the same data after lowering the WaveDeform threshold (modified WaveDeform), and is found to have a SNR\,$=57.9$ and SNR$\,=\,0.69\times10^5$, in the bin with the lowest SNR value and for the full distribution, respectively.
While the noise is observed to increase (as expected) with the modified WaveDeform, the contamination level is still considered minimal (below 2\% in any bin). 




\subsection{Fitting procedure}\label{sec::fitting}
We would now like to fit the charge distribution to extract the SPE charge templates (the components of Eq.~\ref{eq::fit}) for all DOMs. 

Contamination from two-photon events is suppressed by the pulse selection, but can not be entirely avoided. To minimize potential biases by the charge entries resulting from two photons, the one and two photon contribution to the charge distributions is fitted at the same time, using something we call a convolutional fitter. It assumes that the charge distribution resulting from two photons is the SPE charge distribution convolved with itself~\cite{dossi2000methods}. In each step of the minimizer the convolution is updated given the current set of SPE parameters to be evaluated and the relative one and two photon contributions is determined.

We do not fit for the three-photon contribution, which is justified by the lack of statistics in the 3\,PE region as well as the significant rate difference between the 1\,PE and 2\,PE region, as shown in Fig.~\ref{fig::pulse_selection} (right).



Although the pulse selection and the modification to WaveDeform allow us to probe the charge distribution below the discriminator, there is insufficient data to accurately fit Exp$_1$ to each DOM independently. Rather, we determine the value of Exp$_1$ using the aggregate of all data together, and use that result for Exp$_1$ in all subsequent fits.

The motivation to not fit for Exp$_1$ independently for all DOMs is as follows. Pulses that fall below the WaveDeform threshold do not contribute to the pulse series. This is equivalent to an efficiency factor for DOM in question. Having large variations in the Exp$_1$ component will in turn cause large variation in the individual DOMs estimated efficiencies. Rather, setting all DOMs to have the same Exp$_1$ improves the overall fit while reducing the DOM-to-DOM efficiency differences. In IceCube analyses, a global efficiency systematic uncertainty accounts for this. 

Using the aggregate of the entire ensemble of DOMs with the modified WaveDeform datasets, we background-subtract the FT distribution from the Event dataset (removing the electronic noise), and fit the resulting distribution to determine the components of Eq.~\ref{eq::fit}. The determined shape and normalization of Exp$_1$ is then used in all individual DOM fits using the standard WaveDeform.

\begin{wrapfigure}[18]{R}{0.4\textwidth}
\centering
\includegraphics[width=.4\textwidth]{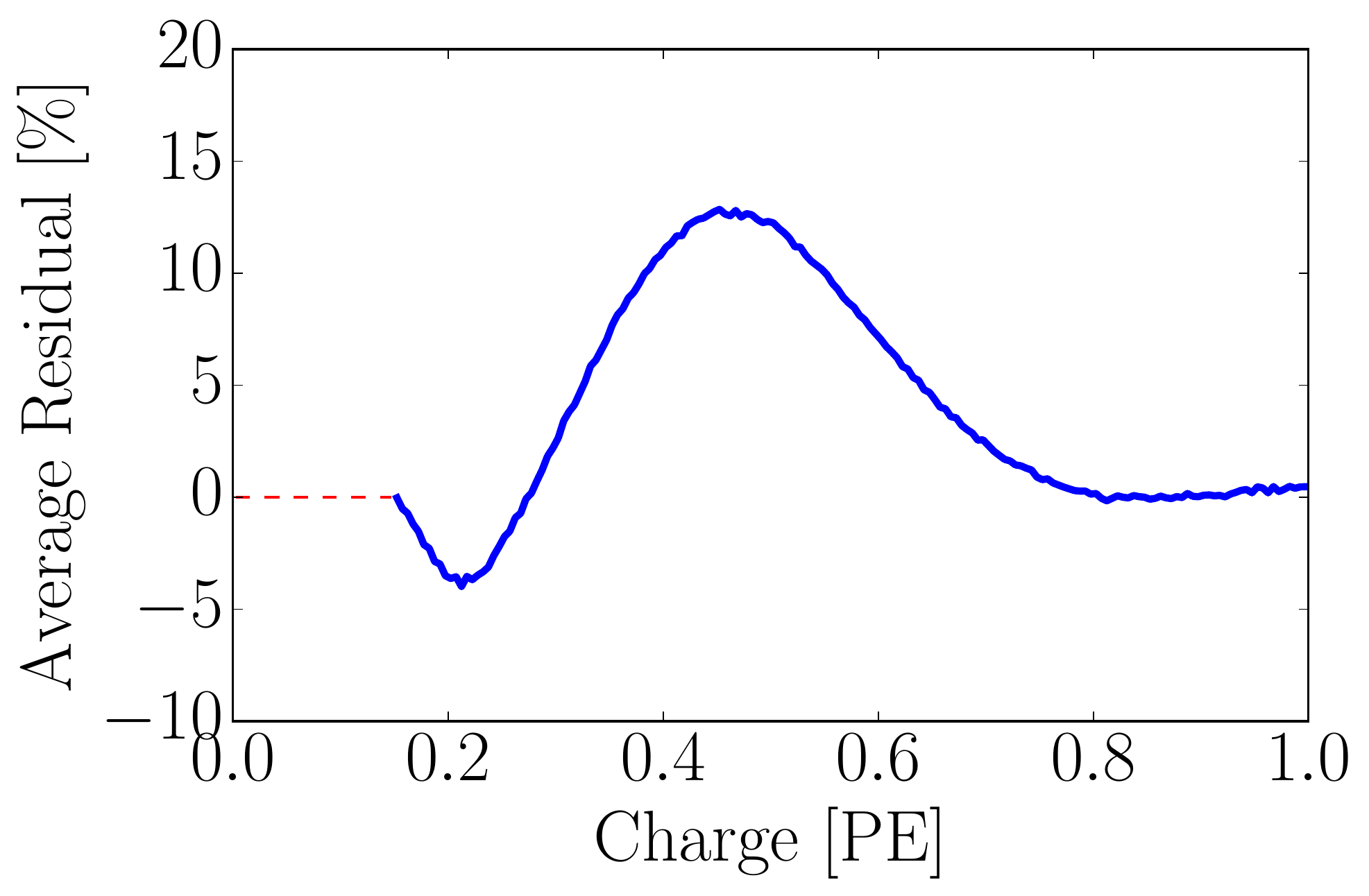}
\caption{The extracted residual correction in blue, comparing the result of the convolutional fit to the data, averaged over all DOMs. The dashed red line indicates the region where we do not have sufficient data and therefore set the residual to 0\% (i.e. no correction will be applied in this region).}
\label{fig::residual}
\end{wrapfigure}

Upon fitting the standard WaveDeform Event dataset with the determined values for Exp$_1$, the residual of each fit is calculated by measuring the percentage difference between the fit and the data. By averaging the residuals over all the DOMs (shown in Fig.~\ref{fig::residual}), a correction function is generated and will be used as a global additive factor for all SPE charge templates to account for the difference between the chosen model (Eq.~\ref{eq::fit}) and the actual data. All SPE charge templates include the residual correction function in Fig.~\ref{fig::residual}.

As described in Sec.~\ref{sec::SPE_Templates}, the Gaussian mean ($\mu$) is used to determine the gain setting for each PMT. Therefore, it is particularly important that the fit quality in the peak region accurately describes the data. While fitting to the full charge distribution improves the overall fit agreement, the mismatch between the chosen functional form (Eq.~\ref{eq::fit}) and a true SPE charge distribution can cause the Gaussian component to pull away from its ideal location. To compensate for this, the fitting algorithm prioritizes fitting to the data around the Gaussian mean. This is accomplished by first fitting to the full distribution to get an estimate of the Gaussian mean location (to within 1\%). Then, the data in the region $\pm$0.15\,PE around the original estimated Gaussian mean is weighted to have a higher impact on the fit, and the distribution is re-fitted. The re-fitting amounts to a percent level change in the Gaussian mean location and the chosen region around the original estimate of the Gaussian mean was found not to produce a significant bias.

\subsection{SPE charge template fit results}\label{sec::fit_results}
We now present the results of the fits and describe the correlations of the fit parameters with hardware differences, and time variations in the next section. 
By subtracting the modified WaveDeform electronic noise spectrum (shown in Fig.~\ref{fig::beacon_launches}) from the Event dataset, the Exp$_1$ component was determined by fitting the aggregate distribution from 0.1~PE to 3.5~PE. The result of the fit yielded P$_{\mathrm{e}1}$~=~0.186\,$\pm$\,0.041 and w$_1$~=~0.027$\,\pm\,$0.002~PE, with a Pearson's correlation coefficient of -0.895. Determining the shape and normalization was the sole purpose of using the modified WaveDeform. All subsequent discussion relates to the standard WaveDeform datasets. The determined values for Exp$_1$ are now used to describe the low-PE charge region for all subsequent standard WaveDeform fits.

Using the Event dataset with the measured values of Exp$_1$, the SPE charge templates are extracted for every DOM, separately for each IceCube season from IC86.2011 to IC86.2016. The fit range for Exp$_2$ and the Gaussian components is selected to be between 0.15\,PE and 3.5\,PE. An average fit was also performed on the charge distribution, in which all the data for a given DOM was summed over all seasons (labeled as "AVG"). 

An SPE charge template for a particular DOM is classified as having a "failed fit" if it is flagged by one of the validity checks. The validity checks ensure that there is sufficient data extracted from the pulse selection and that the fit was successful, as well as goodness of fit check on the reduced $\chi^2$. The majority of the failed fits, approximately 95\%, are due to DOMs which have been removed from service (see Fig.~\ref{fig::DOM_features}). The remaining failed fits were found to be associated with DOMs that have known issues associated with their charge collection. All the DOMs with failed fits are not included in this analysis. In the simulation, the average SPE charge template is assigned to these DOMs. 

We can divide the DOMs into subsets with the following hardware differences: 
\begin{enumerate}
    \item  HQE DOMs with the new toroids,
    \item  Standard QE DOMs with the new toroids,
    \item and Standard QE DOMs with the old toroids.
\end{enumerate}
The mean value and standard error of the mean of the IC86.AVG fit parameters for the subset of hardware differences are listed in Table~\ref{table::fits_avg}. The residual, averaged over all DOMs, from 0 to 1\,PE is shown in Fig.~\ref{fig::residual}. 

\begin{table}[h]
\scriptsize
\begin{center}
\begin{tabular}{||c c c c c c||}
\hline
Configuration & P$_{\mathrm{e}2}$ &  w$_2$  &  N  & $\mu$  &  $\sigma$ \\ [0.5ex]
\hline
\hline
HQE / New Toroid & 0.245 $\pm$ 0.002 & 0.358 $\pm$ 0.004 & 0.569 $\pm$ 0.002 &  1.0163 $\pm$ 0.0011 &  0.311 $\pm$ 0.002 \\
\hline
Std. QE / New Toroids & 0.211 $\pm$ 0.001 & 0.348 $\pm$ 0.001 & 0.602 $\pm$ 0.001 &  1.0198 $\pm$ 0.0004 &  0.315 $\pm$ 0.001 \\
\hline
Std. QE / Old Toroids & 0.207 $\pm$ 0.001 & 0.381 $\pm$ 0.003 & 0.607 $\pm$ 0.001 &  1.0044 $\pm$ 0.0008 &  0.293 $\pm$ 0.001 \\
\hline
\end{tabular}
\caption{The measured mean of each fit parameter and the standard error of the mean for the subset of hardware configurations listed in the first column. The fitted parameters for Exp$_1$, determined at the beginning of this section, are: P$_{\mathrm{e}1}$~=~0.186\,$\pm$\,0.041 and w$_1$~=~0.027$\,\pm\,$0.002~PE. The definitions of each fitted variable can be found in the description of Eq.~\ref{eq::fit}.}
\label{table::fits_avg}
\end{center}
\end{table}

\begin{figure}[h]
\begin{center}
\includegraphics[width=0.75\columnwidth]{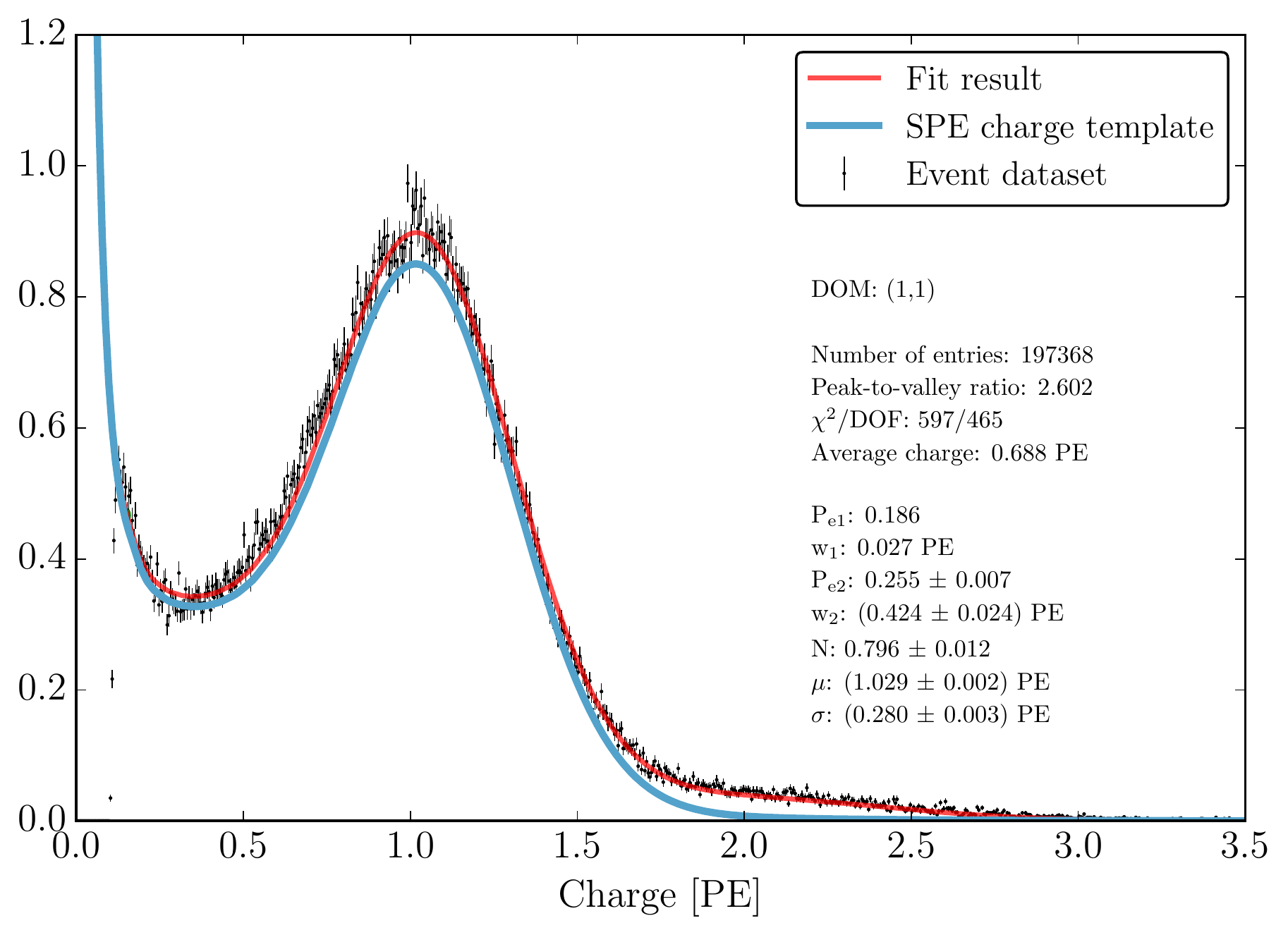}
\caption{The result of the convolutional fit (red) for DOM(1,1) using the Event dataset (black data points with statistical uncertainties), including all data from seasons IC86.2011 to IC86.2016. 
The extracted SPE charge template from the fit result is shown in blue. For both the convolution fit and the SPE charge template, the curves include the correction from the average residual shown in Fig.~\ref{fig::residual}. As discussed in Sec.~\ref{sec::fitting}, P$_{\mathrm{e}1}$~=~0.186  and w$_1$ = 0.027\,PE are fixed for the individual DOM fits. The definitions of the fitted values in the annotation can be found in the description of Eq.~\ref{eq::fit}. The difference between the fit result and the SPE charge template is the 2PE contamination. The 2PE contribution is defined as convolution of the SPE charge template with itself, and the SPE charge template is non-zero at low charges, the 2PE contribution is non-zero at low charges.}
  \label{fig::fit}
 \end{center}
\end{figure}

An example fit is shown in Fig.~\ref{fig::fit} for the Event dataset charge distribution for DOM (1,1). The collected charge distribution is shown in the black histogram, while the fit to the data is shown as the red line. The extracted SPE charge template from the fit is shown in blue. Both the fit result and extracted SPE charge template shown here incorporate the average residual found in Fig.~\ref{fig::residual}.


\section{Discussion}\label{sec::results}

\subsection{Correlations between fit parameters and DOM hardware differences}
\begin{figure}[h]
    \centering
    \begin{minipage}{.48\linewidth}
        \includegraphics[width=\linewidth]{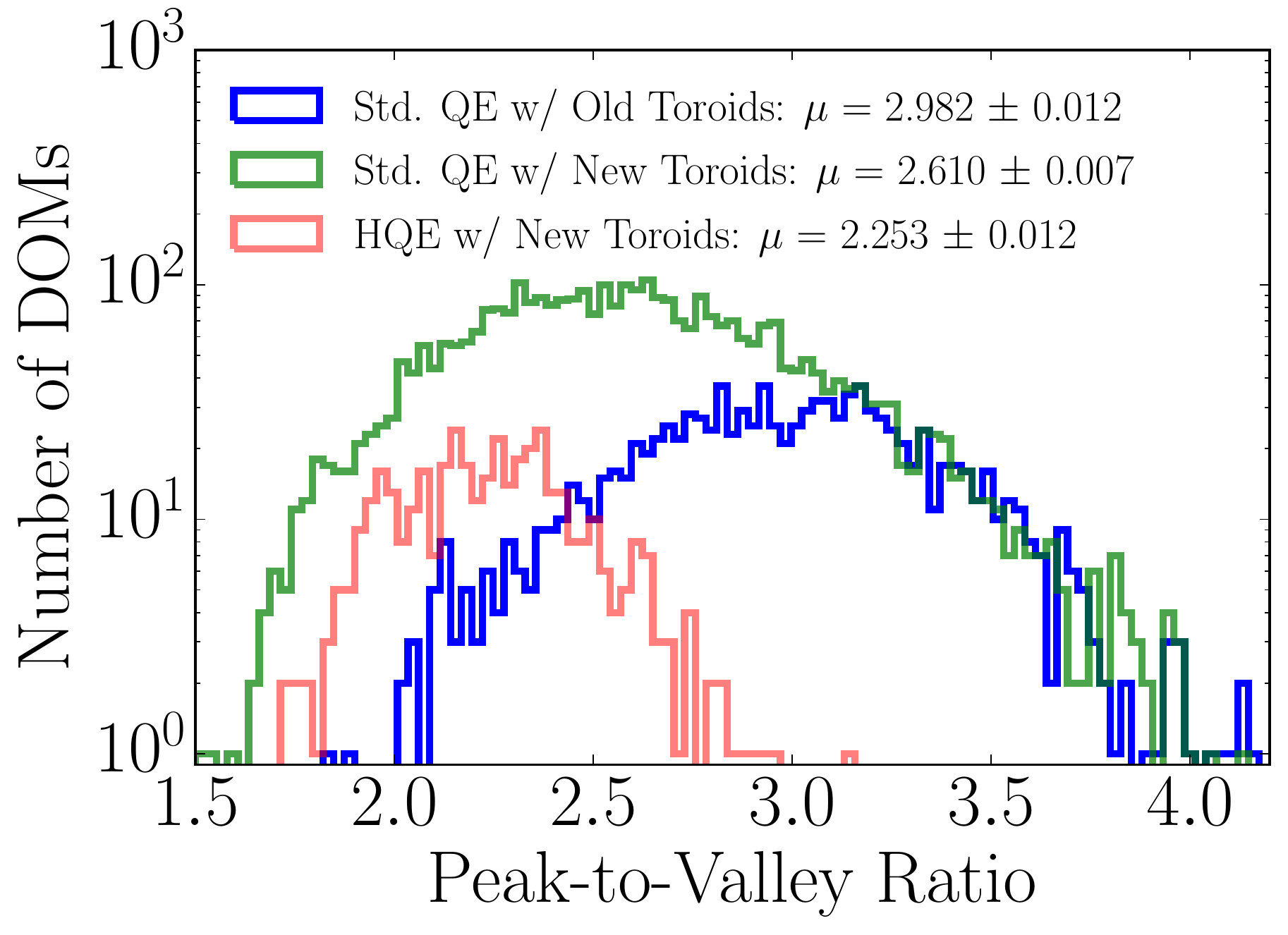}
    \end{minipage}
    \hspace{.01\linewidth}
    \begin{minipage}{.48\linewidth}
        \includegraphics[width=\linewidth]{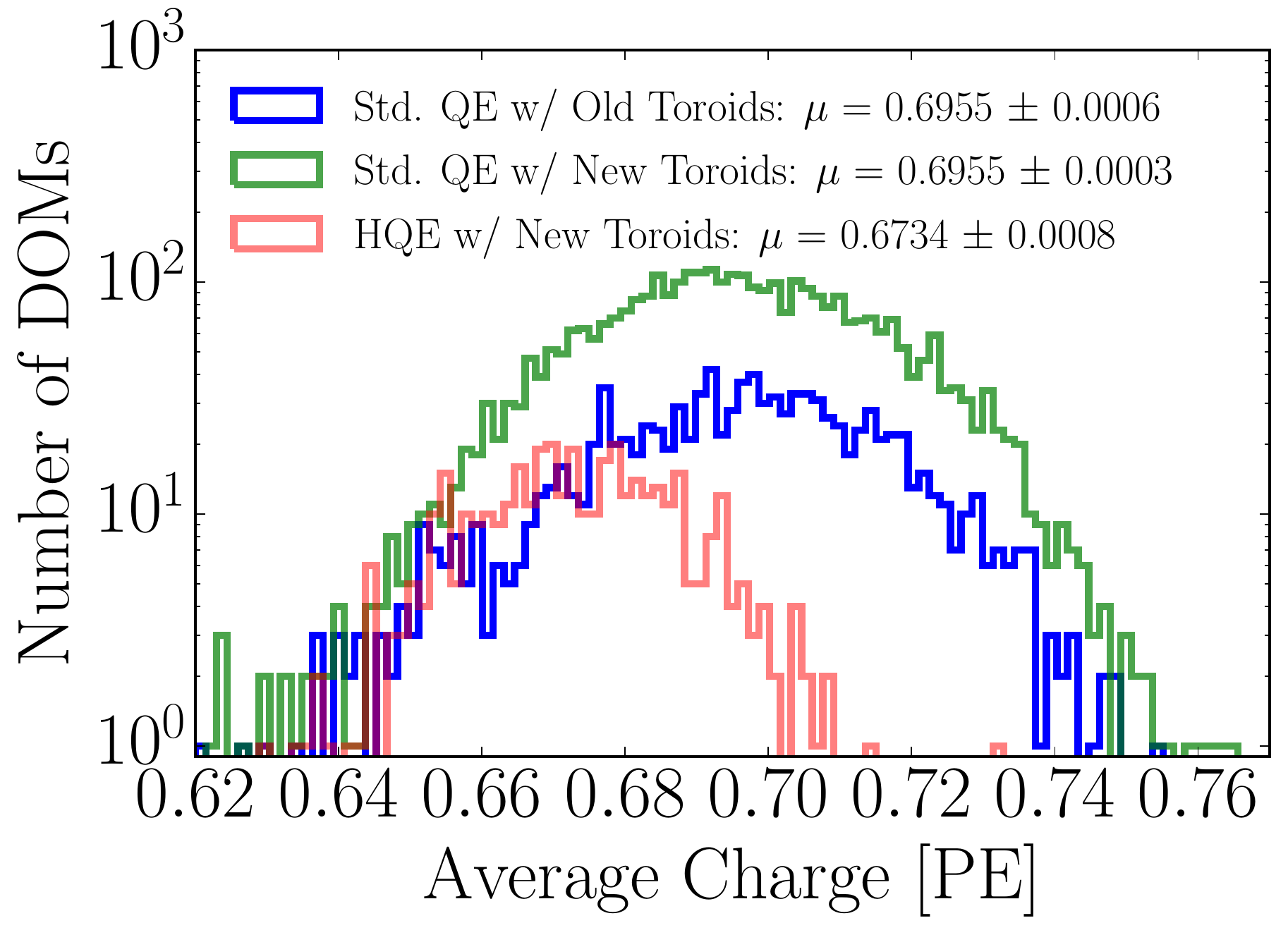}
    \end{minipage}
    \caption{Comparison between the R7081-02MOD HQE DOMs and standard R7081-02 DOMs. Left: The peak-to-valley ratio for the two subsets of quantum efficiencies. Right: The average charge of the individual DOM SPE charge templates. }
    \label{fig::sub}
\end{figure}

It is evident from the data in Table~\ref{table::fits_avg} that the average shape of the SPE charge templates depends on the DOM hardware.These differences can also be seen in the measured peak-to-valley ratios and average charge of the SPE charge template (see Fig.~\ref{fig::sub}). The peak-to-valley ratio is defined as the ratio between the maximum in the SPE charge template near the SPE peak divided by the minimum found near the discriminator. 

When we examine the subset of DOMs instrumented with the new toroids, the average HQE DOM were found to have a 16.1\,$\pm$\,1.1\% larger P$_{\mathrm{e}2}$ component and 5.5\,$\pm$\,0.4\% smaller Gaussian probability. Consequently, the average HQE peak-to-valley ratio is measured to be 2.25\,$\pm$\,0.01, or 13.68\,$\pm$\,0.08\% lower than the average for Standard QE DOMs. Also, the average charge of the average HQE DOM was found to be 3.2\,$\pm$\,0.3\% lower than that of the Standard QE DOMs. The average charge is calculated by integrating over the full SPE charge template including the residual correction. The distribution is shown Fig.~\ref{fig::sub} (right). The average charge is found to be below 1~PE due to the low-PE contribution from Exp$_1$ and Exp$_2$, whose physical description can be found in Sec.~\ref{sec::SPE_Templates}.

IceCube compensates for the change in the mean measured charge in simulation, by increasing the HQE DOM efficiency by the equivalent amount. This ensures that the total amount of charge collected by the HQE DOMs remains the same prior to, and after, inserting the SPE charge templates into simulation. 

Similarly, using only the subset of Standard QE DOMs, the SPE charge templates were found to have measurably different shapes when comparing the different methods of AC coupling. The average Gaussian standard deviation for the DOMs instrumented with the old toroids were found to be  7.0\,$\pm$\,0.7$\%$ narrower, albeit with a very similar probability of the photoelectron populating the Gaussian component. With these differences, we find a peak-to-valley ratio of 2.610\,$\pm$\,0.007 for the new toroid DOMs and 2.982\,$\pm$\,0.012 for the old toroid DOMs, corresponding to a difference of 12.47\,$\pm$\,0.07\%.
The average Gaussian mean of the fit for the DOMs with the old toroids was also found to be 1.5\,$\pm$\,0.2$\%$ lower than those with the new toroids. This corresponds proportionally to a change in the expected gain. The difference in the average charge, however, between these two hardware configurations is very similar. 

\begin{figure}[!h]
\begin{center}
\includegraphics[width=1\columnwidth]{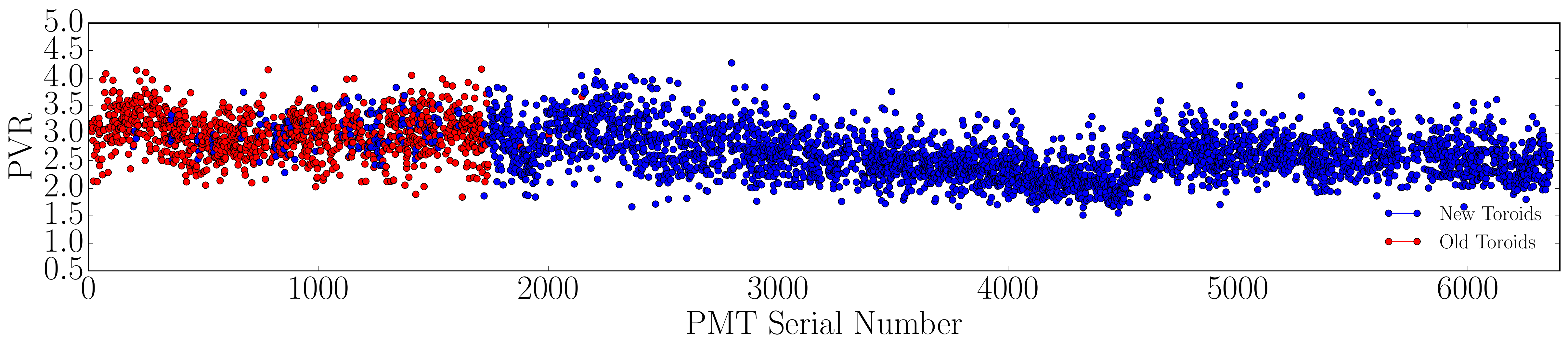}
\caption{The measured peak-to-valley ratio for the standard QE PMTs ordered by PMT serial number. The red data points indicate a PMT instrumented with an old toroid, whereas new toroids are indicated by the blue data points.}
  \label{fig::pvr}
 \end{center}
\end{figure}

Although the DOMs instrumented with the old toroids were deployed into the ice earlier than those with the new toroids, the differences above remain when examining individual deployment years; therefore, the shape differences are not attributed to the change in the DOM behavior over time. However, the DOMs with the old toroids were the first PMTs to be manufactured by Hamamatsu. A gradual change of the fit parameters was observed when ordering the PMTs according to their PMT serial number (i.e. their manufacturing order). Fig.~\ref{fig::pvr} shows the change in the measured peak-to-valley ratio as a function of PMT serial number for the standard QE DOMs with the new toroids (blue) and old toroids  (red). This is compelling evidence that the observed differences between the new and old toroids is due to a change in the PMT production procedure rather than version of AC coupling.

\begin{figure}[t]
\begin{center}
\includegraphics[width=0.7\columnwidth]{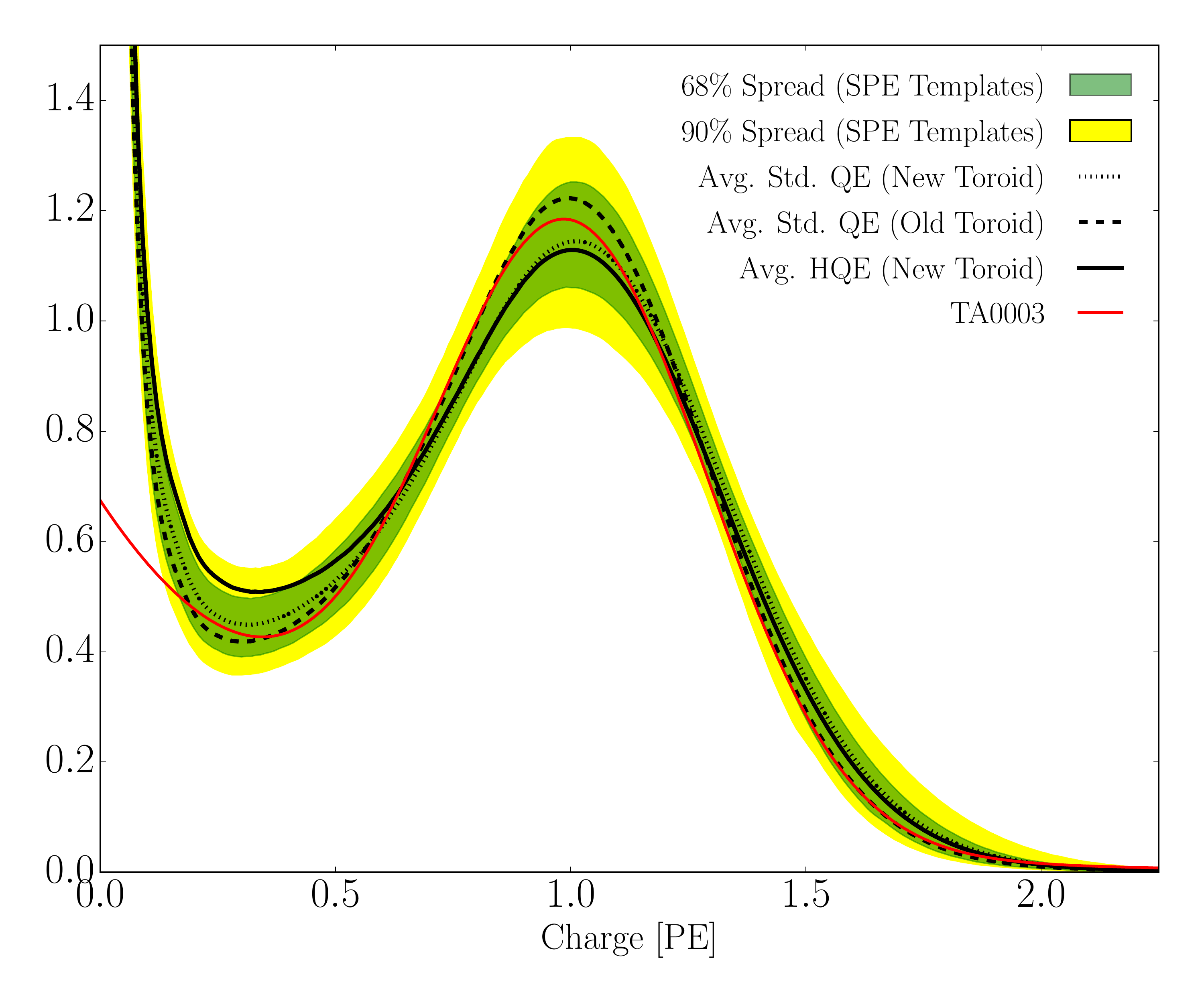}
\caption{The green (yellow) regions show the 68\% (90\%) spread in the SPE charge templates for a given charge.  Superimposed are the average SPE charge templates for the variety of hardware configurations shown in the black dotted, dashed, and solid lines. The TA0003 distribution, for comparison, is shown in red. All curves have been normalized such that the area above 0.23\,PE is the same. 
}
  \label{fig::fits}
 \end{center}
\end{figure}

 Fig.~\ref{fig::fits} illustrates the average shape differences in the extracted SPE charge templates between the HQE DOM with the new toroids (solid black line), Standard QE with the new toroids (dotted black line), Standard QE with the old toroids (dashed black line), compared to the spread in the measured SPE charge templates for all DOMs in the detector (green and yellow intervals). The figure also shows how the previous default SPE charge distribution, the TA0003 distribution, compares to this recent measurement. All curves in this figure have been normalized such that the area above 0.23\,PE is the same. The observable shape differences from the TA0003 are attributed to a better understanding of the low-charge region, the difference in functional form (described in Section~\ref{sec::SPE_Templates}), and the fact that the SPE charge templates were generated using a realistic photocathode illumination.
 

 \subsection{Fitting parameters variation over time}
The SPE charge templates were also extracted for each IceCube season independently (IC86.2011 to IC86.2016), in order to investigate the time dependence of the fit parameters. The data was broken up into IceCube seasons rather than years, since calibration to remove gain drift is performed at the beginning of each IceCube season. 

\begin{figure}[!h]
\begin{center}
\includegraphics[width=0.55\columnwidth]{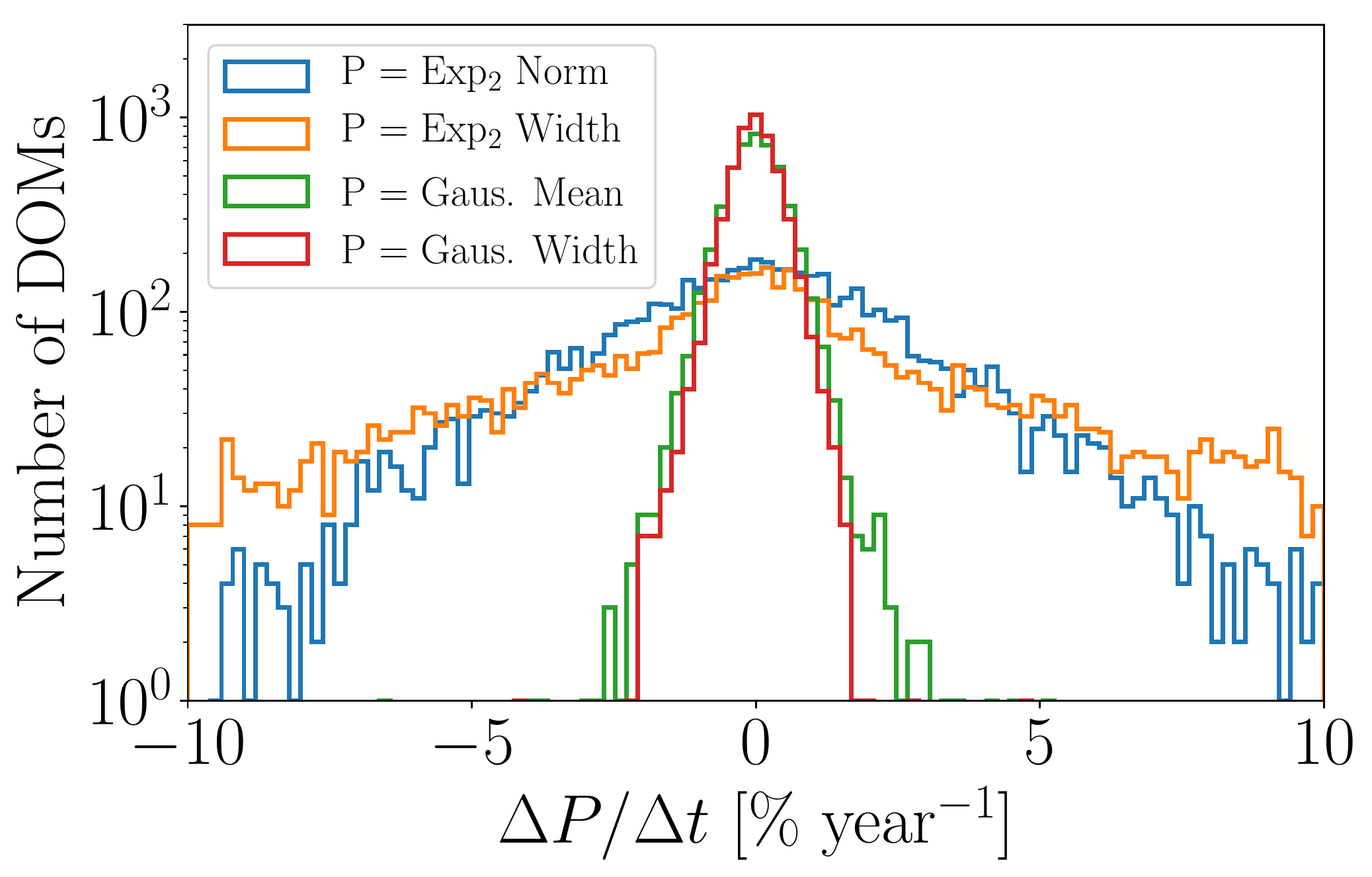}
\caption{The change in the individual DOM fitted parameters, $\Delta$P, over time, $\Delta$t. The histogram represents the rate of change, per IceCube season, of the fitted parameter labelled in the legend. }
  \label{fig::time}
 \end{center}
\end{figure}

For every DOM in the detector, the change over time of each fit parameter (excluding Exp$_1$, since it was held fixed for this analysis) was calculated. Fig.~\ref{fig::time} shows the change in a given fit parameter, relative to the mean value, per IceCube season. The measured distributions were found to be consistent with randomly scrambling the order of each measurement. More than 96\% of DOMs are found to have less than 1.0\% change per year in the measured location of the Gaussian mean, in agreement with Ref.~\cite{icecube_instrumentation}. This observation holds for the individual subsets of DOMs with different hardware configurations as well. 


 
\subsection{Quantifying observable changes when modifying the PMT charge distributions}

In this section we investigate the extent to which changes to the SPE charge templates impact quantities such as the trigger fraction above our discriminator threshold or the measured charge over threshold, since these quantities are used in the analysis of IceCube physics events. If the choice of template has a significant impact on high-level physics results reported by IceCube, it would be a serious concern. Thus we explore the effect of changing the assumed gain response in simulation. The change has different implications depending on the typical illumination level present in different physics analyses. Therefore we explore the effect of the templates in three illumination regimes: dim, semi-bright, and bright illumination.


The PMT response function is described by a combination of a "bare" efficiency, $\eta_0$, and a normalized charge response function, $f(q)$. The bare efficiency represents the fraction of arriving photons that result in any nonzero charge response, including those below the discriminator threshold. The normalization condition is:
\begin{equation}
\int_{0}^{\infty} f(q) dq = 1.
\end{equation}
Generally, $f(q)$ and $\eta_0$ have to be adjusted together to maintain agreement with a quantity known from laboratory or in-ice measurements, such as the predicted number of pulses above threshold for a dim source.

\paragraph*{Dim source measurements}
Where light levels are low enough, the low occupancy ensures that sub-discriminator pulses do not contribute to any observed charge as they do not satisfy the trigger threshold.
Given some independent way of knowing the number of arriving photons, a laboratory or in-ice measurement determines the trigger fraction above threshold $\eta_{0.23}$ and/or the average charge over threshold Q$_{0.23}$, either of which can be used to constrain the model as follows:

\begin{equation}
\eta_{0.23} = \eta_0 \int_{0.23q_{pk}}^{\infty} f(q) \mathrm{d}q
\end{equation}

\begin{equation}
Q_{0.23} = \eta_0 \int_{0.23q_{pk}}^{\infty} q f(q) \mathrm{d}q
\end{equation}

Here, the discriminator threshold is assumed to be 0.23 times the peak charge q$_{pk}$. It is also useful to multiply observed charges by q$_{pk}$, since we set each PMT gain by such a reference, and then a measurement constraint would be stated in terms of Q$_{0.23}$/q$_{pk}$.

\paragraph*{Semi-bright source measurements}
For semi-bright sources, pulses that arrive after the readout time window is opened are not subject to the discriminator threshold. WaveDeform introduces a software termination condition at $\sim$0.13\,PE (described at the end of Section~\ref{sec::pulse_selection}). The average charge of an individual pulse that arrives within the time window is:
\begin{equation}
Q_{0.13} = \eta_0  \int_{0.13q_{pk}}^{\infty} q f(q) \mathrm{d}q 
\end{equation}

\paragraph*{Bright source measurements}
For light levels that are large (e.g. a cosmic ray muon passing near the photocathode), the trigger is satisfied regardless of the response to individual photons, and the total charge per arriving photon therefore includes contributions below both the discriminator and the WaveDeform thresholds:
\begin{equation}
Q_{0} = \eta_0  \int_{0}^{\infty} q f(q) \mathrm{d}q 
\end{equation}

As such, the total charge is directly proportional to the average charge of the SPE charge template.

\subsubsection{Model comparison}\label{sec::model_comparison}
A natural question to ask is whether or not a change in $f(q)$ would cause observable changes in the ratio between dim, semi-bright, and bright source measurements. That is, when we change the SPE charge distribution in simulation, should we expect the charge collected by bright events compared to dim events to change? When the charge distribution model is changed in a way that preserves agreement with the measured $\eta_{0.23}$ or Q$_{0.23}$/q$_{pk}$, i.e.~$\eta_0$ is adjusted properly for changes in $f(q)$, the physical effect can be summarized by the change in in the ratios Q$_0$/Q$_{0.23}$, and Q$_0$/Q$_{0.13}$. Conveniently, these ratios depend only on the shape of $f(q)$. Table~\ref{table::models} compares these ratios in terms of the TA0003 charge distribution and the SPE charge templates described here. It is shown that there are sub-percent level differences in the physically-observable ratios. The largest difference in the shape between the SPE charge templates and the TA0003 distribution is in the low-charge region, particularly below $\sim$0.10\,PE. Charge from this region can only inflate bright events. That is, these pulses are too small to trigger the discriminator or be reconstructed by WaveDeform, but they can reside on top of other pulses, inflating them. Since these pulses by definition contain little charge, they do not tend to inflate the measured charge by a noticeable amount, as shown by the Q$_0$/Q$_{0.23}$ measurements in Table~\ref{table::models}.

\begin{table}[!h]
\footnotesize
\begin{center}
\begin{tabular}{||c c c c ||}
\hline
\textbf{Model} & Detector &$Q_{0}$/$Q_{0.23}$ & $Q_0$/$Q_{0.13}$ \\ [0.5ex]
\hline
\hline

TA0003 & All DOMs &1.014 & 1.005 \\
\hline
SPE Templates & HQE + New Toroids &1.019$\pm$0.002& 1.007$\pm$0.001 \\
\hline
   & Std. QE + New Toroids &1.016$\pm$0.002& 1.006$\pm$0.001  \\
\hline
   & Std. QE + Old Toroids &1.015$\pm$0.002& 1.006$\pm$0.001  \\
\hline
\end{tabular}
\caption{The average value of the ratios for dim, semi-bright, and bright source measurements, for the SPE templates and the previous charge distribution (TA0003).  
The uncertainty reported represents the standard deviation of the distribution of the calculated quantity.}
\label{table::models}
\end{center}
\end{table}

\subsection{SPE charge templates for calibration}

The gain setting on each PMT is calibrated prior to the beginning of each season such that the Gaussian mean of the charge distribution corresponds to a gain of 10$^7$, or equivalently 1\,PE. This gain calibration method, run directly on the DOMs, uses waveform integration for charge determination instead of WaveDeform unfolding, resulting in a small systematic shift in gain. This systematic shift was determined for every PMT. The mean shift obtained over all DOMs was found to be 1.47\,$\pm$\,0.04\% with a standard deviation of 2.62\%, corresponding to an overestimation of the measured charge in the detector. 

The correction to the systematic shift in the measured charge can be implemented retroactively by dividing the reported charge from WaveDeform by the corresponding offset for a given DOM. Alternatively, we can account for this by simply inserting SPE charge templates, measured in this analysis, into simulation such that the corresponding systematic shift is also modeled in simulation. This will be performed in the following subsection.

\subsection{SPE charge templates in simulation}
To model the IceCube instrument, the PMT response is described in the simulation. The IceCube simulation chain assigns a charge to every photoelectron generated at the surface of the photocathode. The charge is determined by sampling from a normalized SPE charge distribution probability density function (PDF). A comparison to data between describing the SPE charge distribution PDF using the SPE charge templates and the TA0003 distribution follows. 

Two Monte Carlo simulation datasets were produced, one using the SPE charge templates to describe the charge distribution for every DOM in the detector and the other using the TA0003 distribution. These two simulation sets were produced using the same events and pass through the same IceCube processing chain as for the final analysis level of the IC86.2011 sterile neutrino analysis~\cite{aartsen2016searches}. Here, the events at the final analysis level are $>$99.9\% upward-going (a trajectory oriented upwards relative to the horizon) secondary muons produced by charged current muon neutrino/antineutrino interactions. The event selection criteria is described in Ref.~\cite{aartsen2015evidence}. The muon reconstructed energy range of this event selection is between approximately 500\,GeV and 10\,TeV.


\begin{figure}[h]
    \centering
        \includegraphics[width=\linewidth]{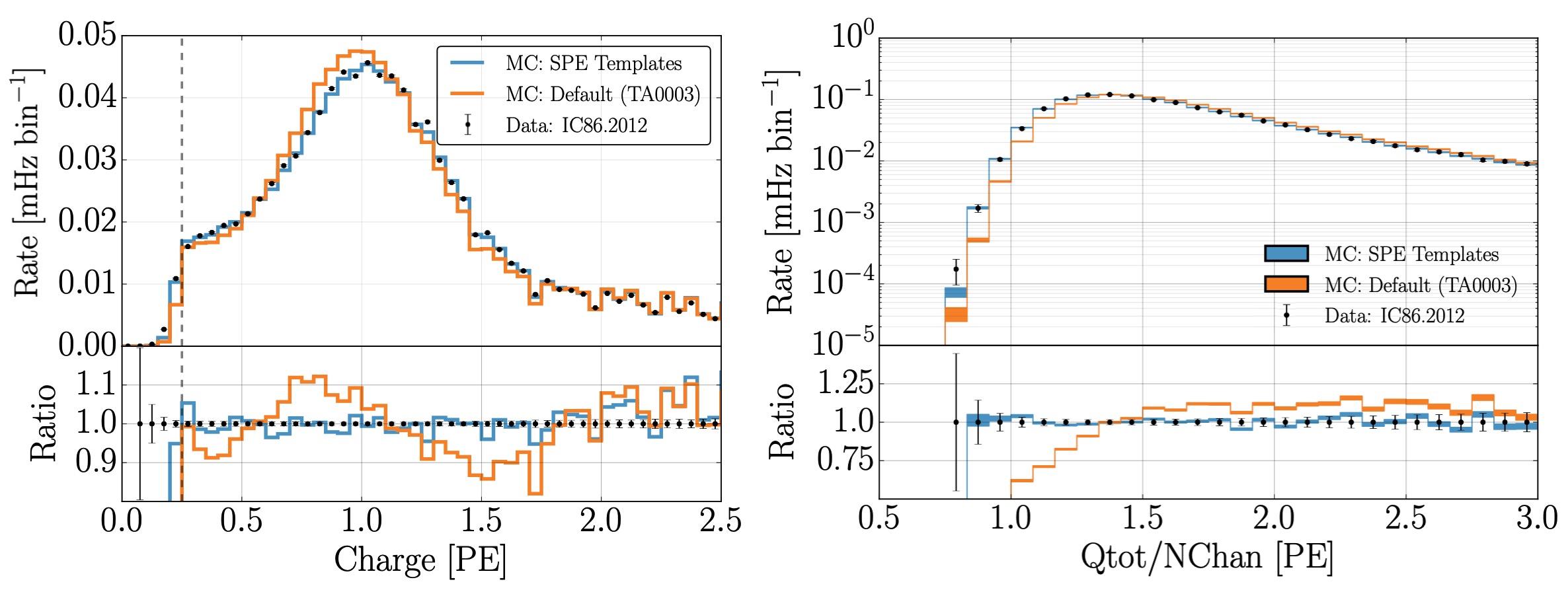}
    \hspace{.01\linewidth}
    \caption{A comparison between IC86.2012 (black data points with statistical-only uncertainties) to two sets of simulation at an IceCube analysis level. Both simulation sets were produced with identical procedures, except for one describing the SPE charge distribution using the SPE charge templates (blue) and the other using the original TA0003 (orange) model. Left: The total measured charge per DOM, per neutrino event. Right: The distribution of the total measured charge (Qtot) per DOM, divided by the number of DOM channels (NChan) that participated in the event.}
\label{fig::mc_test}
\end{figure}

Fig.~\ref{fig::mc_test} (left) shows a histogram of the total measured charge per DOM, during each neutrino interaction, after noise removal. The black data points represent the measured IC86.2012 data but are statistically equivalent to the other seasons. The simulation set using the TA0003 charge distribution is shown in orange, and that using the SPE charge templates is shown in blue.   Fig.~\ref{fig::mc_test} (right) shows the distribution of the total measured charge of an event divided by the number of channels (NChan), or DOMs, that participated in the event. Both plots in Fig.~\ref{fig::mc_test} have been normalized such that the area under the histograms is the same. 

The SPE charge templates clearly improve the overall MC description of these two variables. This update may be useful for analyses that rely on low-occupancy events (low-energy or dim events) in which average charge per channels is below 1.5\,PE, and will be investigated further within IceCube. 

\section{Conclusion}
This article outlines the procedure used to extract the SPE charge templates for all in-ice DOMs in the IceCube detector using in-situ data from IC86.2011 to IC86.2016. The SPE charge templates represent an update to the modeled single photoelectron charge distribution previous used by IceCube, the TA0003 distribution. The result of this measurement was shown to be useful for improving the overall data/MC agreement for an IceCube analysis as well as calibration of the individual PMTs. It also prompted a comparison between the shape of the SPE charge templates for a variety of hardware configurations and time dependent correlations.

The subset of HQE DOMs were found to have a smaller peak-to-valley ratio relative to the Standard QE DOMs, as well as an overall 3.2\,$\pm$\,0.3\% lower average charge. It was also found that the DOMs instrumented with the old toroids used for AC coupling (the first PMTs to be manufactured by Hamamatsu) had narrower Gaussian standard deviation and an increased peak-to-valley ratio of 12.47\,$\pm$\,0.07\%. By assuming a relationship between the production time and serial number, this difference was shown to be likely due to a change in the manufacturing over time rather than the actual AC coupling method. No significant time dependence in any of the fitted parameters associated with the SPE charge templates over the investigated seasons was observed. A reassessment of the PMT gain settings found a systematic bias of 1.47\,$\pm$\,0.04\% with a standard deviation of 2.62\%. 

The SPE charge templates were inserted into the MC simulation and the results were compared to the default TA0003 distribution. A significant improvement in the description of the variables ``total charge per DOM" and ``total charge over the number of channels" was shown.  Analyses which rely on low-light occupancy measurements, will benefit from this update. As shown in the ratios between the dim, semi-bright, and bright source measurements, the average charge for various light levels will not be affected by this update.



\newpage

\acknowledgments
We acknowledge the support from the following agencies: 

USA {\textendash} U.S. National Science Foundation-Office of Polar Programs,
U.S. National Science Foundation-Physics Division,
Wisconsin Alumni Research Foundation,
Center for High Throughput Computing (CHTC) at the University of Wisconsin-Madison,
Open Science Grid (OSG),
Extreme Science and Engineering Discovery Environment (XSEDE),
U.S. Department of Energy-National Energy Research Scientific Computing Center,
Particle astrophysics research computing center at the University of Maryland,
Institute for Cyber-Enabled Research at Michigan State University,
and Astroparticle physics computational facility at Marquette University;
Belgium {\textendash} Funds for Scientific Research (FRS-FNRS and FWO),
FWO Odysseus and Big Science programmes,
and Belgian Federal Science Policy Office (Belspo);
Germany {\textendash} Bundesministerium f{\"u}r Bildung und Forschung (BMBF),
Deutsche Forschungsgemeinschaft (DFG),
Helmholtz Alliance for Astroparticle Physics (HAP),
Initiative and Networking Fund of the Helmholtz Association,
Deutsches Elektronen Synchrotron (DESY),
and High Performance Computing cluster of the RWTH Aachen;
Sweden {\textendash} Swedish Research Council,
Swedish Polar Research Secretariat,
Swedish National Infrastructure for Computing (SNIC),
and Knut and Alice Wallenberg Foundation;
Australia {\textendash} Australian Research Council;
Canada {\textendash} Natural Sciences and Engineering Research Council of Canada,
Calcul Qu{\'e}bec, Compute Ontario, Canada Foundation for Innovation, WestGrid, and Compute Canada;
Denmark {\textendash} Villum Fonden, Danish National Research Foundation (DNRF), Carlsberg Foundation;
New Zealand {\textendash} Marsden Fund;
Japan {\textendash} Japan Society for Promotion of Science (JSPS)
and Institute for Global Prominent Research (IGPR) of Chiba University;
Korea {\textendash} National Research Foundation of Korea (NRF);
Switzerland {\textendash} Swiss National Science Foundation (SNSF);
United Kingdom {\textendash} Department of Physics, University of Oxford.

\newpage
\bibliography{Master}
\bibliographystyle{JHEP}
\newpage

\end{document}